\documentclass[a4paper, 11pt]{article}
\pdfoutput=1
\usepackage{jcappub} 
\usepackage{lineno}
\usepackage{graphicx}
\usepackage{amsmath}
\usepackage{mathtools}
\usepackage{mathrsfs}
\usepackage{microtype}
\usepackage{cancel}
\usepackage{bigints}
\usepackage{bm}
\usepackage[dvipsnames]{xcolor}
\usepackage{soul}
\usepackage{csquotes}
\usepackage{subcaption}
\usepackage{enumitem}
\usepackage{xspace}
\usepackage{booktabs}
\usepackage{siunitx}
\usepackage[ISO]{diffcoeff}
\usepackage{ragged2e}
\usepackage{tensor}
\usepackage[normalem]{ulem}
\usepackage{cleveref}
\usepackage{hyperref}

\makeatletter
\input{aas_macros.sty}
\let\jnl@style=\relax
\makeatother

\DeclareSIUnit{\year}{yr}
\DeclareSIUnit{\pc}{pc}
\DeclareSIUnit{\kpc}{\kilo\pc}
\DeclareSIUnit{\Mpc}{\mega\pc}
\DeclareSIUnit{\cLight}{\text{\ensuremath{c}}}
\DeclareSIUnit{\hHubble}{\text{\ensuremath{h}}}
\DeclareSIUnit{\Msun}{\text{\ensuremath{M_\odot}}}

\newcommand*{\ddelta}{{\mathchar '26\mkern -10mu\delta_{\mathrm{D}}}}
\newcommand{\bx}{{\bm x}}

\newcommand{\bk}{{\bm k}}
\newcommand{\bp}{{\bm p}}
\newcommand{\bq}{{\bm q}}
\newcommand{\bv}{{\bm v}}
\newcommand{\bl}{{\bm l}}
\newcommand{\beq}{\begin{equation}\begin{aligned}\relax}
\newcommand{\eeq}{\end{aligned}\end{equation}}

\newcommand*{\q}{\enquote}
\newcommand*{\ml}{\mleft}
\newcommand*{\mr}{\mright}
\DeclareMathOperator{\FFT}{FFT}
\newcommand{\code}{\textsc}
\newcommand*{\expect}[1]{\mleft\langle #1 \mright\rangle}
{\makeatletter
    \intertext@
    \global\let\nobreakintertext=\intertext
\makeatother}
\newcount\dummycount
\ExplSyntaxOn
\patchcmd{\nobreakintertext}{\penalty}{\dummycount}{}{}
\patchcmd{\nobreakintertext}{\penalty}{\dummycount}{}{}
\ExplSyntaxOff

\definecolor{green2}{cmyk}{0.27, 0, 1, 0.52}
\hypersetup{
    colorlinks,            
    linkcolor=green2,      
    citecolor=green2,      
    filecolor=magenta,     
    urlcolor=green2        
}

\title{Early Growth of Structure in\texorpdfstring{\\}{ }Warm Wave Dark Matter}
\author[a]{Mustafa A.\ Amin,}
\emailAdd{mustafa.a.amin@rice.edu}
\author[b]{Simon May,}
\emailAdd{simon.may@pitp.ca}
\author[c]{Mehrdad Mirbabayi}
\emailAdd{mirbabayi@ictp.it}

\affiliation[a]{Department of Physics and Astronomy, Rice University, Houston, TX, 77005, U.S.A.}
\affiliation[b]{Perimeter Institute for Theoretical Physics, Waterloo, ON, N2L 2Y5, Canada}
\affiliation[c]{International Centre for Theoretical Physics, Trieste, Italy}

\abstract{We explore the growth of structure in wave-like dark matter models, where the field and density spectra are peaked at sub-horizon wavenumbers. Starting with the Schrödinger–Poisson system, we derive the scale-dependent evolution of the matter power spectrum during radiation and matter domination. We find a suppression of adiabatic perturbations during radiation domination, controlled by a free-streaming length, and scale-dependent growth of the initially white-noise isocurvature power, controlled by a Jeans scale during matter domination. The results are in qualitative, and in some regimes quantitative, agreement with the quasi-particle picture. We verify the analytic results of the power spectrum with $3+1$-dimensional cosmological Schrödinger–Poisson simulations. We propose an analytic formula for the halo mass function, which is in rough agreement with the simulation results at early times after matter–radiation equality. Our simulations show that early halos typically host a soliton.}

\begin{document}

\maketitle
\flushbottom

\section{Introduction}
\label{sec:intro}
The microscopic nature of dark matter (DM) and its production mechanism are not yet known \cite{Cirelli:2024ssz}. We know that it must be massive and interact predominantly via gravitation. If it is made of bosonic particles, the mass can range from \SI{e-19}{\eV} \cite{Dalal, Amin:2022nlh} to \SI{e19}{\GeV}, assuming a single fundamental species, with an even broader range when allowing for multiple species or composite objects \cite{ParticleDataGroup:2020ssz}. For masses $\lesssim \si{\eV}$, the large occupation numbers allow for a ``classical" field treatment of DM \cite{Marsh:2015xka, Hui:2016ltb, Ferreira:2020fam}. The quantum chromodynamics (QCD) axion \cite{Preskill:1982cy}, axion-like particles \cite{OHare:2024nmr}, vector or dark-photon dark matter \cite{Antypas:2022asj}, tensor dark matter \cite{Marzola:2017lbt} (or more generally spin-$s$ bosonic fields \cite{Jain:2021pnk}) etc.\ are all amenable to such a treatment.

The initial conditions for such wave-like matter can be set \emph{during} inflation, or via local production mechanisms \emph{after} inflation. Due to causality considerations, post-inflationary mechanisms lead to fields and corresponding density perturbations with significant variations on horizon and sub-horizon scales. An example of such a production mechanism is axion dark matter from a Peccei–Quinn phase transition after inflation. See \cite{Buschmann:2021sdq, Gorghetto, Saikawa:2024bta} and references therein.\footnote{In contrast with the post-inflationary scenario, inflationary production of light scalars leads to an essentially homogeneous field. Inflationary production of light vector fields, however, cannot produce such a homogeneous field; the density is dominated by sub-horizon field modes \cite{Graham, Kolb:2020fwh} similar to the post-inflationary cases \cite{Agrawal, Dror:2018pdh, Long:2019lwl, Adshead:2023qiw, Cyncynates:2023zwj}.}

As a result, post-inflationary production leads to an enhanced isocurvature density power spectrum on small scales. Moreover, when dark matter is extremely light, these inhomogeneities result in a non-negligible velocity dispersion and hence free-streaming suppression of the adiabatic spectrum \cite{Amin:2022nlh}. The lack of observation of these effects in Ly$\alpha$ data was used to put a lower bound on the mass of post-inflationary dark matter in \cite{McQuinn, Amin:2022nlh, Long:2024imw}. To reduce theoretical uncertainties in the above bound, and more importantly, provide accurate predictions for potential future detections, it is necessary to calculate the shape and time dependence of the matter power spectrum (and associated observables) in detail. That is the goal of the present paper. Since some of us have been involved in very recent related work \cite{Ling:2024qfv, Amin:2025dtd}, we review connections to them first and highlight what is new in the present paper. 

Post-inflationary production leads to an initial field configuration which can be thought of as a dynamical collection of sub-horizon de Broglie scale patches of size $a/k_* < 1/H$. Approximating these patches as quasi-particles with mass $\sim \bar{\rho}(a/k_*)^3$ (where $\bar{\rho}$ is the mean dark matter density), the isocurvature enhancement and free-streaming/Jeans suppression can be understood from their finite number density and the velocity dispersion. An analytical derivation based on solving a truncated Bogoliubov–Born–Green–Kirkwood–Yvon (BBGKY) hierarchy of the scale-dependent time evolution of the density power spectrum was provided by some of us in \cite{Amin:2025dtd}. These results matched well with results from $N$-body simulations, both at the level of the power spectrum in the linear regime, and the estimated halo mass function in the nonlinear regime. This quasi-particle picture was expected to hold for $k < k_*$.

In the present work, we include effects at $k \sim k_*$, as well as some additional wave-dynamical effects on scales $k \ll k_*$. Instead of starting with the BBGKY hierarchy, we start from the Schrödinger–Poisson system, and arrive at evolution equations for the matter power spectrum. These are similar to the corresponding equations in \cite{Amin:2025dtd}, but not identical. For $k \ll k_*$, these effects do not change the quasi-particle results qualitatively. The present formalism also allows us to make some limited predictions regarding the formation of halos, and solitons, which are distinctly wave-dynamical phenomena.

To verify and go beyond our analytic derivation, we carry out $3+1$-dimensional cosmological simulations of the Schrödinger–Poisson system. We find excellent agreement in the overlapping domains of validity between numerical results and analytical estimates of this paper, as well as those of \cite{Amin:2025dtd}. Our simulations differ from usual fuzzy dark matter simulations, which assume a homogeneous mode of the field \cite{Schive:2014dra}. One of the novel aspects of the simulations here is the setup of appropriate isocurvature and adiabatic initial conditions (ICs) for fields with nontrivial spectra. We use the results in \cite{Ling:2024qfv}, with some modifications. In \cite{Ling:2024qfv}, results of relativistic field simulations, along with initial conditions were provided in the radiation-dominated era. However, unlike the present paper they were not extended to the matter-dominated era where the self-gravity of dark matter is relevant.

Before describing the organization of this paper, we briefly review the relevance of our results for current and upcoming observations. A variety of these observations aim to probe the small-scale matter power spectrum \cite{Drlica-Wagner:2022lbd}. These include analyses of the Ly$\alpha$ forest, galaxy satellite populations, gravitational lensing, stellar streams, \SI{21}{\cm} line intensity mapping, dynamical heating of stars, and the abundance of high-redshift galaxies
\cite{Mondino:2020rkn, Sabti:2021unj, Gilman:2021gkj, Delos:2021ouc, Drlica-Wagner:2022lbd, Boylan-Kolchin:2022kae, Chung:2023syw, Irsic:2023equ, Delos:2023dwq, Esteban:2023xpk, Nadler:2024ims, Xiao:2024qay, Ji:2024ott, deKruijf:2024voc, Lazare:2024uvj, Buckley:2025zgh, He:2025jwp}. Among the small-scale features, the ``free-streaming suppression" has been extensively studied in microscopic particle dark matter models \cite{Narayanan:2000tp, Hansen:2001zv, Lewis:2002nc, Green:2003un, Viel:2005qj, Lesgourgues:2006nd, Viel:2007mv, Boyarsky:2008xj, Erickcek:2011us, Lancaster:2017ksf, Irsic:2017ixq, wdm, Miller:2019pss, Erickcek:2021fsu, Ballesteros:2020adh, Sarkar:2021pqh, Garcia:2023qab}, and more recently in wave dark matter models \cite{Amin:2022nlh, Liu:2024pjg, Ling:2024qfv, Long:2024imw, Amin:2025dtd, Liu:2025lts}. Similarly, the small-scale enhancement and their implications have been studied in, for example \cite{Dai:2019lud, Ramani:2020hdo, Lee:2020wfn, Blinov:2021axd, Lee:2021zqw, Delos:2021rqs, Gilman:2021gkj, Esteban:2023xpk, Graham:2024hah}.

For warm wave dark matter, the free streaming-, white noise-, and Jeans-related features in the DM density spectrum can be accessible on quasi-linear scales ($k\sim \SI{10}{\per\Mpc}$). Our calculation of the linear power spectrum in warm wave dark matter (as well as the mass function) can be useful in making predictions for these systems and potentially detecting or constraining dark matter properties and production mechanisms. By using the linear matter power spectrum as input, we can go deeper into the nonlinear regime by estimating the mass function, as well as the formation times of halos and solitons.

The rest of the paper has the following structure. We provide an analytic derivation of the evolution of the density power spectrum in \cref{sec:Derivation}. In \cref{sec:Numerics}, we discuss the numerical setup, including initial conditions for the field and its time evolution. \Cref{sec:SimResults} contains simulation results in the linear and nonlinear regime, and comparisons with analytic results. We end with a discussion and summary in \cref{sec:summary}. We follow the metric and Fourier conventions in \cite{Amin:2025dtd}, which are provided in \cref{sec:notconv} for convenience. \Cref{sec:summaryPS} provides a summary of the results for the power spectrum evolution in terms of convenient dimensionless variables (used in many of our figures). \Cref{sec:examples} provides explicit formulae related to specific initial field spectra which we used for demonstrating the agreement between analytic and numerical results. In \cref{sec:NumericalDetails}, we provide details of how to set up initial conditions, the numerical algorithm for the time evolution of the fields, as well as resolution considerations and numerical convergence checks.

\section{Derivation}
\label{sec:Derivation}
Starting with the Schrödinger–Poisson system in an expanding universe, we derive the time evolution of the density power spectrum. The initial field configuration may be relativistic. However with sufficient time, due to redshifting of the field gradients, we  eventually end up with the field configuration being predominantly non-relativistic. We take such a time when the field is non-relativistic to be the initial time for our analytic (and numerical) calculations. This initial time is much earlier than matter–radiation equality since the field is supposed to be dark matter. We also restrict ourselves to sub-horizon scales in what follows.\footnote{For systematic treatments of relativistic corrections to the non-relativistic equations in the linear and nonlinear regimes of structure formation in wave dark matter, see for example \cite{Salehian:2020bon, Salehian:2021khb}.}
All lengths and coordinates are taken to be comoving unless stated otherwise.

\subsection{Schrödinger–Poisson System}
A self-gravitating, non-relativistic scalar field $\varphi(t,\bx) = \Re[\Psi(t,\bx)e^{-imt}]/\sqrt{2m}$ with mass $m$ in an expanding universe evolves according to the Schrödinger–Poisson system of coupled equations. Using time and field variables defined via $\dl\eta /\dl t = 1/a^2$ and $\psi = a^{3/2}\Psi$, with the scale factor $a(t)$, the equations read%
\begin{gather}
    \label{eq:SPscalar}
    i\partial_\eta \psi = -\frac{\nabla^2}{2m}\psi + m a^2 \Phi \psi ,
    \qquad
    \nabla^2\Phi = 4\pi G a^2 \ml(\rho - \bar{\rho}\mr),
    \nobreakintertext{where the ``\emph{physical}" density is}
    \rho(\eta,\bx) = m|\Psi(\eta,\bx)|^2 = m\frac{|\psi(\eta,\bx)|^2}{a^3},
\end{gather}%
with $\bar{\rho}$ denoting its spatial average, $G$ is Newton's constant, and $\Phi(\eta,\bx)$ is the gravitational potential.

Let $c_\bq^+$, $c_\bq^-$ be the Fourier transforms of $\psi$ and $\psi^*$. Then,
\beq
    \label{eq:Dcpm}
    &\pm i\partial_\eta c^{\pm}_\bp = \frac{p^2}{2m} c_\bp + m a^2 \int_\bq\Phi_\bq c^{\pm}_{\bp-\bq},
\eeq
where we used the shorthand $\int_\bq = \int \dl\bq/(2\pi)^3$. We assume $\Phi_{\bm 0} = 0$, and for $\Phi_{\bq\ne\bm{0}}$, the Poisson equation yields
\beq
    \Phi_\bq = -\frac{4\pi Ga^2}{q^2}\rho_\bq \,,
    \quad \text{with} \qquad
    \rho_\bq = \frac{m}{a^3}\int_\bp c_\bp^+ c_{\bq-\bp}^- \,.
\eeq
In the post-inflationary scenario, $c_\bp^\pm$ are random variables. A natural choice for their \emph{initial} statistics is
\beq
    \label{eq:cstat}
    &\langle c_\bp^+(\eta_0) c_\bq^-(\eta_0)\rangle =
    \frac{\bar{\rho}a^3}{m} f_0(p)\ddelta(\bp + \bq) ,
    \quad \text{with} \quad
    \int_\bp f_0(p) = 1 ,
\eeq
where $\ddelta(\bk) = (2\pi)^3\delta_{\mathrm{D}}(\bk)$.
Note that $f_0(p)$ is essentially the power spectrum of the field $\psi$:
\beq
    f_0(p) = \frac{m}{a^3\bar{\rho}} P_\psi(p),
\eeq
and also the one-particle phase space distribution function.\footnote{This can be seen by evaluating the kinetic energy density of the Schrödinger field $\Psi = a^{3/2}\psi$:
\beq
    \rho_{\mathrm{K}} = -\frac{1}{2m a^2}\expect{\Psi^*\nabla^2 \Psi}= \frac{\bar{\rho}}{m}\int_\bp \frac{p^2}{2m a^2}f_0(p).
    \nonumber
\eeq
}
For concreteness, we will assume that the field power spectrum is dominated by a comoving momentum $k_*$, which corresponds to the characteristic de Broglie scale at production.

To obtain the evolution of the matter density, it is convenient to define the bilinear \cite{Levkov:2018kau}
\beq
    \label{eq:bilinearf}
    \hat{f}_\bk(\bp) \equiv \frac{m}{a^3\bar{\rho}}c^+_{\bp+\bk/2}c^-_{-\bp+\bk/2}\,,
\eeq
in terms of which we can write the fractional density perturbation and the gravitational potential, as
\beq
    \label{eq:deltaPhi}
    \delta_\bk = \frac{\rho_\bk}{\bar{\rho}}=\int_\bp \hat{f}_\bk(\bp),\qquad \Phi_\bk=-\frac{4\pi G\bar{\rho}a^2}{k^2}\int_\bp \hat{f}_\bk(\bp).
\eeq

\subsection{Adiabatic Perturbations}
\label{sec:AdPert}
Given a realization of adiabatic perturbations, the perturbed Fourier coefficients of the field can be computed by a matching procedure.
The density and velocity perturbations $\delta_{\mathrm{ad}}(\eta_0, \bx)$ and $\theta_{\mathrm{ad}}(\eta_0, \bx)$ shift the initial field $\psi$ (generated in absence of adiabatic perturbations $\zeta$) as follows:
\beq
    \label{eq:psi-ic}
    \psi(\eta_0, \bx) \rightarrow \psi(\eta_0, \bx) \sqrt{1 + \delta_{\mathrm{ad}}(\eta_0, \bx)} \, e^{i S_{\mathrm{ad}}(\eta_0, \bx)} ,
\eeq
with $\nabla^2 S_{\mathrm{ad}} = m\theta_{\mathrm{ad}}$.
At an early enough time $\eta_0$ before free-streaming effects are important but after the dark matter can be treated as a non-relativistic field that interacts only gravitationally, a linear approximation in the density and velocity perturbations yields \cite{Amin:2022nlh}
\beq
    \label{eq:cin}
    c^\pm_\bp(\eta_0) \to c^\pm_\bp(\eta_0) + \frac{1}{2} \int_\bk \delta_\bk^{\mathrm{ad}}(\eta_0) c^\pm_{\bp-\bk} \mp \int_\bk\theta_\bk^{\mathrm{ad}}(\eta_0) \frac{im}{k^2} c^\pm_{\bp-\bk} ,
\eeq
Deep in the radiation era, $\delta^{\mathrm{ad}}_\bk(\eta_0) \approx 6\zeta_\bk (a_{\mathrm{eq}} k_{\mathrm{eq}}/\sqrt{2})(\eta_0 - \eta_k)$ and $\theta_\bk^{\mathrm{ad}}(\eta_0) = -\partial_{\eta_0} \delta^{\mathrm{ad}}_\bk(\eta_0)$, where $\zeta_\bk$ is the standard adiabatic mode and $\eta_k$ is the time at which $k \approx 1.6a H$ \cite{Dodelson:2020bqr, Baumann:2022mni}. These are good approximations for $k \gg k_{\mathrm{eq}}$. The more accurate relation between $\delta^{\mathrm{ad}}_\bk$ and $\zeta_\bk$ for lower $k$ can be found in \cref{sec:AlgoIC}.

After the shift in \cref{eq:cin}, $f_{\bk}(\bp) = \big\langle\hat f_\bk(\bp)\big\rangle \neq 0$ but proportional to $\delta_\bk^{\mathrm{ad}}(\eta_0)$ and $\theta_\bk^{\mathrm{ad}}(\eta_0)$, which we temporarily treat as independent (cf.~\cref{eq:Pad}). The average ``$\langle\hdots\rangle$" here is over initial field configurations. The time evolution of $f$ can be written as:
\beq
    f_{\bk}(\eta,\bp) = \delta_\bk^{\mathrm{ad}}(\eta_0) \gamma^{(\mathrm{a})}_\bk(\eta, \eta_0, \bp) - \theta_\bk^{\mathrm{ad}}(\eta_0)\gamma_\bk^{(\mathrm{b})}(\eta,\eta_0,\bp),
\eeq
with the initial conditions
\beq
    \label{eq:gamma-ic-ab}
    &\gamma^{(\mathrm{a})}_\bk (\eta_0, \eta_0, \bp) = \frac{f_0(|\bp-\bk/2|)+f_0(|\bp+\bk/2|)}{2} ,
    \\
    &\gamma^{(\mathrm{b})}_\bk (\eta_0, \eta_0, \bp) = \Delta_\bk f_0(p) \equiv (-im/k^2) \left[f_0(|\bp+\bk/2|)-f_0(|\bp-\bk/2|)\right],
\eeq
These follow from substituting \cref{eq:cin} in \cref{eq:bilinearf} and taking an expectation value using \cref{eq:cstat}.

We can use \cref{eq:Dcpm} to derive an equation for $\partial_\eta \hat{f}_\bk$. Taking its ensemble average over initial field configurations, and linearizing in $\delta_\bk^{\mathrm{ad}}$ and $\theta_\bk^{\mathrm{ad}}$ gives the evolution equation for $\gamma^{(\mathrm{a}, \mathrm{b})}_\bk$:
\beq
    \left(\partial_\eta + i\frac{\bp\cdot\bk}{m} \right)\gamma^{(i)}_{\bk}(\bp)+\frac{3\bar{H}_0^2a}{2} \Delta_{\bk} f_0(\bp)\int_\bl \gamma^{(i)}_{\bk}(\bl) \approx 0 .
\eeq
where $\bar{H}_0^2 = (8\pi G/3)\bar{\rho}a^3$. Equivalently,
\beq
    \label{eq:gilbert}
    &\gamma^{(i)}_{\bk}(\eta,\eta_0,\bp) =
    \\
    &\qquad \gamma^{(i)}_\bk (\eta_0, \eta_0,\bp)e^{-i\frac{\bp\cdot \bk}{m}(\eta-\eta_0)} +
        \frac{3\bar{H}_0^2}{2} \int_{\eta_0}^\eta \dl\eta' a(\eta') e^{-i\frac{\bp\cdot \bk}{m}(\eta-\eta')} \Delta_{\bk}f_0(\bp) \int_\bl \gamma_{\bk}(\eta',\eta_0,\bl),
\eeq
which is a close analog of the Gilbert equation \cite{Brandenberger:1987}. Integrating over the momentum allows us to obtain the transfer functions,
\beq
    \label{eq:Ti}
    T^{(i)}_k(\eta,\eta_0) = \int_\bp \gamma^{(i)}_\bk(\eta,\eta_0,\bp) ,
\eeq
which satisfy
\beq
    \label{eq:Tab}
    T_k^{(i)}(\eta,\eta_0) &= T^{\mathrm{fs}\,(i)}_k(\eta,\eta_0)
    + \frac{3\bar{H}_0^2}{2}\int_{\eta_0}^\eta \dl\eta' a(\eta')T^{\mathrm{fs}\,(\mathrm{b})}_k(\eta,\eta') T_k^{(i)}(\eta',\eta_0),
\eeq
with the following free-streaming kernels
\beq
    \label{eq:Tfsab}
    T^{\mathrm{fs}\,(\mathrm{a})}_k(\eta,\eta_0) &= \cos\ml[\frac{k^2}{2m}(\eta-\eta_0)\mr]\int_\bp f_0(p)e^{-i\frac{\bp\cdot\bk}{m}(\eta-\eta_0)},
    \\
    T^{\mathrm{fs}\,(\mathrm{b})}_k(\eta,\eta_0) &= \frac{2m}{k^2} \sin\ml[\frac{k^2}{2m}(\eta-\eta_0)\mr]\underbrace{\int_\bp f_0(p)e^{-i\frac{\bp\cdot \bk}{m}(\eta-\eta_0)}}_{\equiv T^{\mathrm{fs}}_k}.
\eeq
The free-streaming kernels are transfer functions for density and velocity perturbations respectively in absence of gravity. The $T^{(\mathrm{a}, \mathrm{b})}_k$ are transfer functions including both free-streaming and gravitational clustering effects: $\delta_{\bk}(\eta) = \delta_{\bk}(\eta_0) T^{(\mathrm{a})}_k(\eta, \eta_0)$, $\theta_{\bk}(\eta) = \theta_{\bk}(\eta_0) T^{(\mathrm{b})}_k(\eta, \eta_0)$. $T^{\mathrm{fs}}_k$ (without an ``$(\mathrm{a}, \mathrm{b})$" superscript) is the free-streaming kernel used in \cite{Amin:2025dtd}, where we worked in the quasi-particle picture.

Expressing $\delta_\bk^{\mathrm{ad}}$ and $\theta_\bk^{\mathrm{ad}}$ in terms of $\zeta_\bk$ results in\footnote{Note that $P_\delta(\eta,k)\ddelta(\bk_1+\bk_2)=\langle\langle\delta_{\bk_1}\delta_{\bk_2}\rangle\rangle \approx \langle\langle\delta_{\bk_1}\rangle \langle\delta_{\bk_2}\rangle\rangle = P_\zeta(k)\ddelta(\bk_1+\bk_2) [T^{(\mathrm{ad})}_k(\eta,\eta_0)]^2$, where the inner $\langle \delta_\bk\rangle$ is an ensemble average of the initial $c^\pm_\bp$ in a particular realization of $\zeta$, whereas the outer $\langle \dots \rangle$ is an ensemble average over $\zeta$.}
\beq
    \label{eq:Pad}
    P_{\delta}(\eta,k) \supset P_{\zeta}(k) \left[T^{(\mathrm{ad})}_k(\eta,\eta_0)\right]^2,
\eeq
where $T^{(\mathrm{ad})}_k(\eta,\eta_0)$ satisfies the same \cref{eq:Tab} (``$i = \mathrm{ad}$") with
\beq
    \label{eq:Tfsad}
    T_k^{\mathrm{fs}\,(\mathrm{ad})}(\eta,\eta_0)
    &=6\frac{a_{\mathrm{eq}} k_{\mathrm{eq}}}{\sqrt{2}} \left[(\eta_0-\eta_k) T^{\mathrm{fs}\,(\mathrm{a})}_k(\eta, \eta_0) + T^{\mathrm{fs}\,(\mathrm{b})}_k(\eta, \eta_0)\right],
    \\
    &\approx 6\frac{a_{\mathrm{eq}} k_{\mathrm{eq}}}{\sqrt{2}}\frac{2m}{k^2} \sin\ml[\frac{k^2}{2m}(\eta-\eta_k)\mr] T^{\mathrm{fs}}_k(\eta,\eta_0),
\eeq
where we used $(k^2/2m)(\eta_0-\eta_k) \ll 1$ in the last equality.

\subsection{Isocurvature Perturbations}
Instead of solving for the evolution of $\delta_\bk$ in a realization of $\zeta_\bk$, we could have first taken the average over the initial adiabatic perturbations. This would restore translation symmetry, yielding $f_\bk(\bp) = 0$ for $\bk \neq {\bm 0}$, and encode the information about the adiabatic power in the connected, and hence non-Gaussian, 4-point correlations of $c^\pm_\bp$. Solving for the evolution of this connected correlator is an \emph{alternative} way to obtain the same expression as \cref{eq:Pad}. However, it is the \emph{unique} way to understand the growth of the Poisson noise resulting from the randomness of field fluctuations. So we will study it next.

By using the evolution equations for $c^{\pm}_\bp$, we can obtain the evolution equation for the expectation values $\langle \hat{f}_{\bk_1}(\bp_1)\hat{f}_{\bk_2}(\bp_2)\rangle$, which will in turn yield the evolution of the power spectrum of density
\beq
    P_\delta(k_1) \ddelta(\bk_1 + \bk_2) =
    \int_{\bp_1,\bp_2} \langle\hat{f}_{\bk_1}(\bp_1)\hat{f}_{\bk_2}(\bp_2)\rangle ,
\eeq
where we assumed statistical homogeneity and isotropy. The resulting evolution equation for $\langle\hat{f}_{\bk_1}(\bp_2) \hat{f}_{\bk_2}(\bp_2)\rangle$ is:
\beq
    &i\partial_\eta \langle\hat{f}_{\bk_1}(\bp_1)\hat{f}_{\bk_2}(\bp_2)\rangle =
    \\
    &\qquad \frac{\bp_1 \cdot \bk_1}{m}\langle\hat{f}_{\bk_1}(\bp_1) \hat{f}_{\bk_2}(\bp_2)\rangle + (1 \leftrightarrow 2) - {}
    \\
    &\qquad \frac{3}{2}\bar{H}_0^2 a\int_{\bq,\bl} \frac{m}{q^2} \left\langle\hat{f}_\bq(\bl) \left[\hat{f}_{\bk_1-\bq}\ml((\bp_1-\bq)/2\mr) - \hat{f}_{\bk_1-\bq}\ml((\bp_1 + \bq)/2\mr)\right] \hat{f}_{\bk_2}(\bp_2)\right\rangle + (1 \leftrightarrow 2),
\eeq
where we used \cref{eq:bilinearf,eq:Dcpm,eq:deltaPhi}.

The correlation function has a disconnected (Gaussian) and a connected (non-Gaussian) part:
\beq
    \langle\hat{f}_{\bk_1}(\bp_1)\hat{f}_{\bk_2}(\bp_2)\rangle = \underbrace{g_{\bk_1\bk_2}(\bp_1,\bp_2)}_{\text{connected}}+\underbrace{f_0(|\bp_1+\bk_1/2|)f_0(|\bp_2-\bk_1/2|)\ddelta(\bp_1-\bp_2)\ddelta(\bk_1+\bk_2)}_{\text{disconnected}}.
\eeq
Averaging over the initial adiabatic fluctuations results in a nontrivial initial condition for $g$. However, even if the initial condition is trivial ($g=0$), gravitational interactions will generate such non-Gaussianities as we will see momentarily.

We first obtain the density power spectrum due to the disconnected part. From statistical homogeneity and isotropy, $g_{\bk_1\bk_2}(\bp_1,\bp_2)=\tilde{g}_{k_1}(\bp_1,\bp_2)\ddelta(\bk_1+\bk_2)$. Then, the equation for $f_0(p)$ can be obtained in terms of $\tilde{g}_{k}$ via $c^\pm_\bp$ evolution equations (see \cref{eq:Dcpm}) as
\beq
    \partial_\eta f_0(p) = \frac{3\bar{H}_0^2}{2}a \int_\bk \Delta_\bk \tilde{g}_k(\bk,\bp),
\eeq
where $\Delta_\bk \tilde{g}_k(\bk,\bp) = -im/k^2 \ml[\tilde{g}_k(\bk,\bp+\bk/2)-\tilde{g}_k(\bk,\bp-\bk/2)\mr]$. We will ignore this time variation of $f_0$ due to $g$ when using $f_0$ in the equation for $g$ below. This is a good approximation as long as density perturbation evolution is linear.\footnote{See the discussion in section 2.3.2 of \cite{Amin:2025dtd}, with $1/\bar{n}$ there replaced by the initial $P_{\delta}^{(\mathrm{iso})}$ in \cref{eq:Piso_initial}.} The density power spectrum for this time-independent disconnected part is obtained by integrating $\langle \hat{f}_{\bk_1} \hat{f}_{\bk_2}\rangle$ over $\bp_1\bp_2$:
\beq
    \label{eq:Piso_initial}
    P_{\delta}(\eta,k) \supset P_{\delta}^{(\mathrm{iso})}(\eta_0,k)=\int_\bp f_0(|\bp-\bk/2|)f_0(|\bp+\bk/2|),
\eeq
which has an approximately white spectrum at low $k$.

Using the equation for the evolution of the correlation function, as well as the split into connected and disconnected part, the connected part satisfies an equation of the form $\mathcal{D}g = S$:
\beq
    \label{eq:Dg=S}
    &\partial_\eta g_{\bk_1\bk_2}(\bp_1,\bp_2)+\frac{i}{m}(\bp_1\cdot\bk_1+\bp_2\cdot\bk_2)g_{\bk_1\bk_2}(\bp_1,\bp_2) + {}
    \\
    &\qquad\frac{3}{2}\bar{H}_0^2a \left[\Delta_{\bk_1}f_0(\bp_1)\int_\bl g_{\bk_1\bk_2}(\bl,\bp_2)+\Delta_{\bk_2}f_0(\bp_2)\int_\bl g_{\bk_1\bk_2}(\bp_1,\bl)\right]\\
    &\approx -\frac{3}{2}\bar{H}_0^2a \left[\Delta_{\bk_1}f_0(\bp_1)f_0(\bp_2+\bk_2/2|)f_0(|\bp_2-\bk_2/2|)\ddelta(\bk_1+\bk_2)+(1\leftrightarrow2)\right].
\eeq
It is worth noting that in the {quasi-}particle case discussed in \cite{Amin:2025dtd}, $\Delta_\bk f_0(p)\rightarrow -im\bk/k^2 \cdot \nabla_p f_0(p)$. A general solution to the evolution equation for $g_{\bk_1\bk_2}$ is
\beq
    \label{eq:Dg=S-solution}
    &g_{\bk_1\bk_2}(\bp_1, \bp_2) =
    \\
    &\quad \underbrace{\sum_{lm} \gamma^{(l)}_{\bk_1}(\bp_1)\gamma^{(m)}_{\bk_2}(\bp_2) \ddelta(\bk_1+\bk_2)}_{\text{source-free solution}}
    + \underbrace{\frac{3\bar{H}_0^2}{2}\int_{\eta_0}^{\eta} \!\dl\eta' a(\eta') \gamma^{(\mathrm{b})}_{\bk_1}(\bp_1) \gamma^{(\mathrm{c})}_{\bk_2}(\bp_2)\ddelta(\bk_1+\bk_2) + (1 \leftrightarrow 2) \vphantom{\sum_{lm}}}_{\text{sourced solution}},
\eeq
provided that each of $\gamma^{(i)}_{\bk}$ satisfy \cref{eq:gilbert}. We can use the source-free part to get the correct adiabatic power (see the procedure in \cite{Amin:2025dtd}). We have already obtained the adiabatic power spectrum via a different route in \cref{sec:AdPert}, so we have no need to carry out that exercise here.

The sourced solution, on the other hand, gives the gravitational clustering of the isocurvature perturbations. The $\gamma^{(\mathrm{b})}_\bk$ and $\gamma^{(\mathrm{c})}_\bk$ appearing in this solution must have initial conditions given respectively in \cref{eq:gamma-ic-ab} and
\beq
    \label{eq:gamma-ic-c}
    \gamma^{(\mathrm{c})}_\bk(\eta_0, \eta_0, \bp) = f_0(|\bp-\bk/2|) f_0(|\bp+\bk/2|) .
\eeq
This can be seen by constructing a retarded Green's function and making sure it has the correct initial conditions based on the source. The resulting transfer functions $T^{(\mathrm{b},\mathrm{c})}_k$, defined via \cref{eq:Ti}, satisfy the same \cref{eq:Tab} where for $T^{(\mathrm{c})}_k$ we define a new free-streaming kernel
\beq
    \label{eq:Tfsc}
    T^{\mathrm{fs}\,(\mathrm{c})}_k(\eta,\eta_0) = \int_\bp f_0(|\bp-\bk/2|) f_0(|\bp+\bk/2|) e^{-i\frac{\bp \cdot \bk}{m}(\eta - \eta_0)}.
\eeq
The connected part of the density power spectrum $\int_{\bp_1,\bp_2}g_{\bk_1\bk_2}(\bp_1,\bp_2)$ obtained from the sourced solution is then
\beq
    \label{eq:PdIso}
    P_{\delta}(\eta,k) \supset 3\bar{H}_0^2\int_{\eta_0}^\eta \dl\eta' a(\eta') T_k^{(\mathrm{b})}(\eta,\eta')T_k^{(\mathrm{c})}(\eta,\eta').
\eeq

\subsection{Main Result}
We arrive at our main result by putting together the adiabatic part (\cref{eq:Tfsad}), and the isocurvature part, which includes a disconnected (\cref{eq:Piso_initial}) and a connected (\cref{eq:PdIso}) piece,
\beq
    \label{eq:final}
    P_{\delta}(\eta, k)
    = P_{\zeta}(k) \ml[T^{(\mathrm{ad})}_k(\eta,\eta_0)\mr]^2 + P_{\delta}^{(\mathrm{iso})}(\eta_0, k) \left[1 + 3\bar{H}_0^2\int_{\eta_0}^\eta \dl\eta' a(\eta') T_k^{(\mathrm{b})}(\eta,\eta') \hat{T}_k^{(\mathrm{c})}(\eta,\eta')\right],
\eeq
where we defined the normalized transfer function $\hat{T}^{(\mathrm{c})}_k = T^{(\mathrm{c})}_k/P_\delta^{(\mathrm{iso})}(\eta_0,k)$. $T^{(\mathrm{ad},\mathrm{b},\mathrm{c})}_k$ can be obtained from \cref{eq:Tab} with free-streaming kernels in \cref{eq:Tfsab,eq:Tfsad,eq:Tfsc}.\footnote{%
    In practice, it is enough to solve the integral (Volterra) \cref{eq:Tab} only for $T^{(\mathrm{b})}$. Denoting the integral operation in \cref{eq:Tab} by $\star$, the equation takes the form
    $
        T^{(i)} = T^{\mathrm{fs}\,(i)} + T^{\mathrm{fs}\,(\mathrm{b})} \star T^{(i)}.
    $
    We can $\star$ the $(i)$ equation from the right by $T^{\mathrm{fs}\,(\mathrm{b})}$ and the $(\mathrm{b})$ equation from the left by $T^{\mathrm{fs}\,(i)}$ to prove $T^{\mathrm{fs}\,(\mathrm{b})} \star T^{(i)} = T^{(\mathrm{b})} \star T^{\mathrm{fs}\,(i)}$. As a result
    \begin{equation*}
        T^{(i)} = T^{\mathrm{fs}\,(i)} + T^{(\mathrm{b})} \star T^{\mathrm{fs}\,(i)} ,
    \end{equation*}
    which is an explicit expression for $T^{(i)}$, $i \neq \mathrm{b}$, given $T^{(\mathrm{b})}$. See \cite{Amin:2025dtd} for more details.%
}
In the following, we provide an overview of the key physical effects determining the scale-dependent evolution of the power spectrum, and relevant scales.

\subsubsection{Evolution with Free Streaming and Jeans Scales}
The second term on the RHS of \cref{eq:Tab} is responsible for gravitational clustering. At early times, during the radiation era, it is negligible and we recover the same result found in \cite{Amin:2022nlh}. In particular, $T^{(\mathrm{ad})} \sim T^{\mathrm{fs}\,(\mathrm{ad})}$ suppresses the power above the free-streaming wavenumber $k > k_{\mathrm{fs}}(\eta)$, where
\beq
    \label{eq:kfs}
    k_{\mathrm{fs}}(\eta) = \left[\int_{\eta_0}^\eta \dl\eta'a(\eta') \sigma(\eta')\right]^{-1} ,
    \qquad \text{with} \quad
    \sigma(\eta) \equiv \frac{1}{am}\left(\frac{1}{3}\int_\bp p^2f_0(p)\right)^{1/2}
    \sim \frac{k_*}{am} ,
\eeq
In the last equality of the $\sigma(\eta)$ definition, $p^3f_0(p)$ is assumed to have a dominant peak at $k_*$. The free-streaming length is the comoving distance traveled by waves from $\eta_0$ to $\eta$. It is dominated by the motion during the radiation era, saturating to
\beq
    k_{\mathrm{fs}}(\eta > \eta_{\mathrm{eq}}) \sim \frac{a_{\mathrm{eq}}k_{\mathrm{eq}} m}{k_* \ln(m/H_{\mathrm{eq}})}
\eeq
in the matter era, where $H_{\mathrm{eq}}$ is the Hubble parameter at matter–radiation equality. Physically, free-streaming erases existing correlations, such as the adiabatic part of the spectrum.

\begin{figure}
    \centering
    \includegraphics[width=0.75\linewidth]{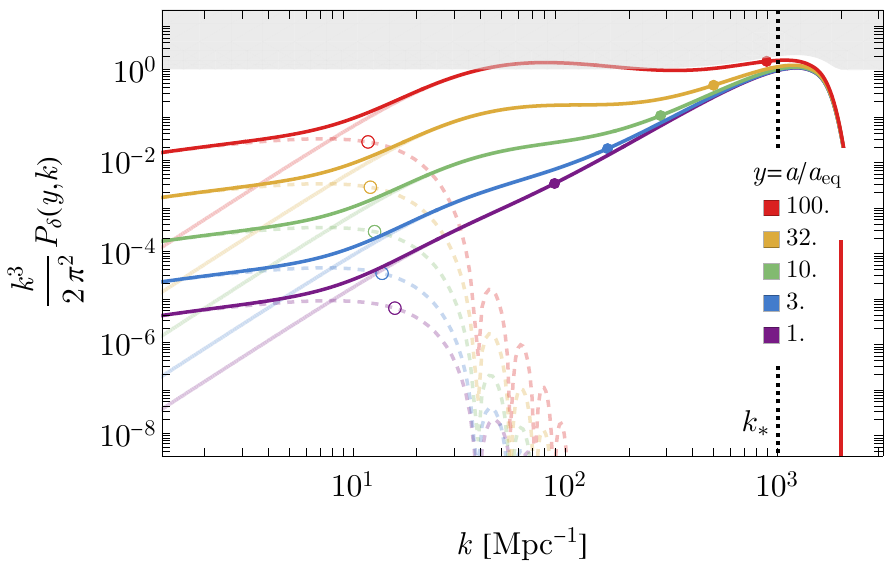}

    \caption{Evolution of the  sum of the adiabatic and isocurvature density contrast power spectrum (darker solid lines) for $m = \SI{e-19}{\eV}$ and $k_*= \SI{e3}{\per\Mpc} = \SI{6.4e-27}{\eV}$, consistent with \cite{Amin:2022nlh}, and $f_0(q) \propto \Theta(k_*-q)$. The open circles indicate the free-streaming wavenumber $k_{\mathrm{fs}}(y)$, whereas solid circles are the Jeans wavenumber $k_{\mathrm{J}}(y)$ where $y=a/a_{\mathrm{eq}}$. The lighter dashed lines are the adiabatic part, which includes the free-streaming suppression for $k > k_{\mathrm{fs}}(y=1)$. The lighter solid lines show the isocurvature part with the usual $y^2$ growth for $k < k_{\mathrm{J}}(y=1) \equiv k_{\mathrm{J}}^{\mathrm{eq}}$ and suppressed growth for $k > k_{\mathrm{J}}^{\mathrm{eq}}$. The gray region delineates the domain of validity of our derivation: $k^3/2\pi^2 [P_\delta(y,k) - P_\delta^{(\mathrm{iso})}(y_0,k)] < 1$.}
    \label{fig:plotPSAdIso}
\end{figure}

During matter domination, the second term on the RHS of \cref{eq:Tab} grows $\propto a(t)$ iff
\beq
    \label{eq:kJ}
    k < k_{\mathrm{J}}(\eta) \equiv a(\eta)\frac{\sqrt{4\pi G\bar{\rho}(\eta)}}{\sigma(\eta)} \sim \sqrt{\frac{a(\eta)}{a_{\mathrm{eq}}}} \frac{a_{\mathrm{eq}}k_{\mathrm{eq}} m}{k_*}.
\eeq
The comoving Jeans length $k_{\mathrm{J}}(\eta)^{-1}$ provides a measure of the distance waves travel over a Hubble time $H(\eta)^{-1}$. However, since $k_{\mathrm{J}}(\eta)\gg k_{\mathrm{fs}}$ (at least by a factor of $10$ at $\eta_{\mathrm{eq}}$) adiabatic perturbations that do not suffer from free-streaming suppression have essentially the same growth factor as in CDM. On the other hand, the initially white-noise isocurvature power is deformed by a scale-dependent growth
\beq
    \label{eq:PisoScalings}
    \frac{P^{(\mathrm{iso})}_\delta (\eta,k)}{P^{(\mathrm{iso})}_\delta(\eta_0,k)} \sim
    \begin{cases}
        \left(\frac{a(\eta)}{a_{\mathrm{eq}}}\right)^2, &\quad
        k < k_{\mathrm{J}}(\eta_{\mathrm{eq}})
        \\
        \left(\frac{k_{\mathrm{J}}(\eta)}{k}\right)^4, &\quad
        k_{\mathrm{J}}(\eta_{\mathrm{eq}}) \leq  k < k_{\mathrm{J}}(\eta)
        \\
        \quad 1, &\quad
        k_{\mathrm{J}}(\eta) \leq k
    \end{cases}
    \,.
\eeq
The simplified behavior in the intermediate $k$-regime is only valid late in the matter-dominated era, with a shallower $k$-dependence at earlier times. Our expression for the power spectrum evolution, \cref{eq:PdIso}, correctly captures these scale-dependent features in the power spectrum similar to what we saw in \cite{Amin:2025dtd}, but with additional refinement for wave dynamics on $k\sim k_*$; see \cref{fig:plotPSAdIso}.

Before highlighting the differences between the wave and quasi-particle pictures, we comment on two recent papers with some overlap with the present work: They explore how interference effects in warm wave dark matter affect the growth driven by non-gravitational \cite{Capanelli:2025nrj} and gravitational \cite{Liu:2025lts} interactions. While the underlying principles are common with us, the distinction between the two important scales $k_{\mathrm{fs}}$ and $k_{\mathrm{J}}$ and their redshift dependence is not very transparent in these works because their analysis is done in a non-expanding setup. There is a semantic difference as well between us and \cite{Liu:2025lts}: Our free-streaming length corresponds to the maximal free-streaming length they refer to in their discussion, whereas our Jeans length corresponds to their local free-streaming length.

Secondly, because gravitational interactions have been treated perturbatively in \cite{Liu:2025lts}, they can only recover a brief initial stage of gravitational growth. In contrast, our analysis, even though perturbative in $\delta$, is non-perturbative in gravitational interactions, which results in a large growth ($\sim 10^3$ before $\delta$ becomes nonlinear).

\subsubsection{Wave Effects}
\label{sec:waveeffects}
Comparing the final result, \cref{eq:final}, with that obtained using the quasi-particle picture in \cite{Amin:2025dtd}, we can identify two wave-dynamical features.

The first wave-dynamical feature is due to the appearance of the terms $\cos(\Delta\eta k^2/(2m))$ and $(2m/k^2) \sin(\Delta\eta k^2/(2m))$ in the expressions for $T^{\mathrm{fs}\,(\mathrm{a}, \mathrm{b})}_k$ (see \cref{eq:Tfsab}). In the quasi-particle picture of \cite{Amin:2025dtd}, these were unity and $\Delta\eta$ respectively. The wavenumber at which these cosines and sines start changing appreciably within a Hubble time is $k_{\mathrm{j}}(\eta) = a(\eta)\sqrt{mH(\eta)}$. This wavenumber corresponds to the ``fuzzy" Jeans scale commonly used in the literature \cite{Hu:2000ke}, but it differs from our Jeans scale $k_{\mathrm{J}}$ defined in \cref{eq:kJ} based on the momentum distribution of the field:\footnote{Heuristically, $k_{\mathrm{J}}$ can be determined by balancing the kinetic and potential terms in the Schrödinger equation $k_*^2/(2m) \sim a^2 m G\bar{\rho}/k_{\mathrm{J}}^2 \Longrightarrow k_{\mathrm{J}} \sim a\sqrt{G\rho}/(k_*/(am)) \sim a^2mH/k_*$, where we used $\Phi\sim G\bar{\rho}\delta/k^2$ and assume an order-unity density perturbation. We can use a similar argument to determine $k_{\mathrm{j}}$: $k_{\mathrm{j}}^2/(2m) \sim a^2 m G\bar{\rho}/k_{\mathrm{j}}^2 \Longrightarrow k_{\mathrm{j}} \sim a\sqrt{m\sqrt{G\bar{\rho}}}\sim a\sqrt{m H}$.}
\beq
    k_{\mathrm{j}}(\eta) \sim \sqrt{k_* k_{\mathrm{J}}(\eta)}.
\eeq
When $k_{\mathrm{J}}(\eta)/k_{\mathrm{j}}(\eta)$ is small, velocity dispersion plays a significant role in structure formation while in the opposite regime the wave effects are predominant. The quasi-particle limit of \cite{Amin:2025dtd} corresponds to the former case, where the ratio is made arbitrarily small by sending $k_*\to \infty$ with $k_*/(a_{\mathrm{eq}} m)$ fixed. Hence, by properly normalizing the arguments of the sine and cosine in the free-streaming kernels (see \cref{sec:summaryPS}), we define
\beq
    \label{eq:gamma}
    \gamma \equiv
    \frac{1}{2\sqrt{2}}\frac{a_{\mathrm{eq}} m}{k_*} \frac{k_{\mathrm{eq}}}{k_*} \sim
    \frac{k_{\mathrm{J}}^2(\eta_{\mathrm{eq}})}{k_{\mathrm{j}}^2(\eta_{\mathrm{eq}})}
\eeq
as a measure of such wave-dynamical effects.

\begin{figure}[t]
    \centering
    \includegraphics[width=1.0\linewidth]{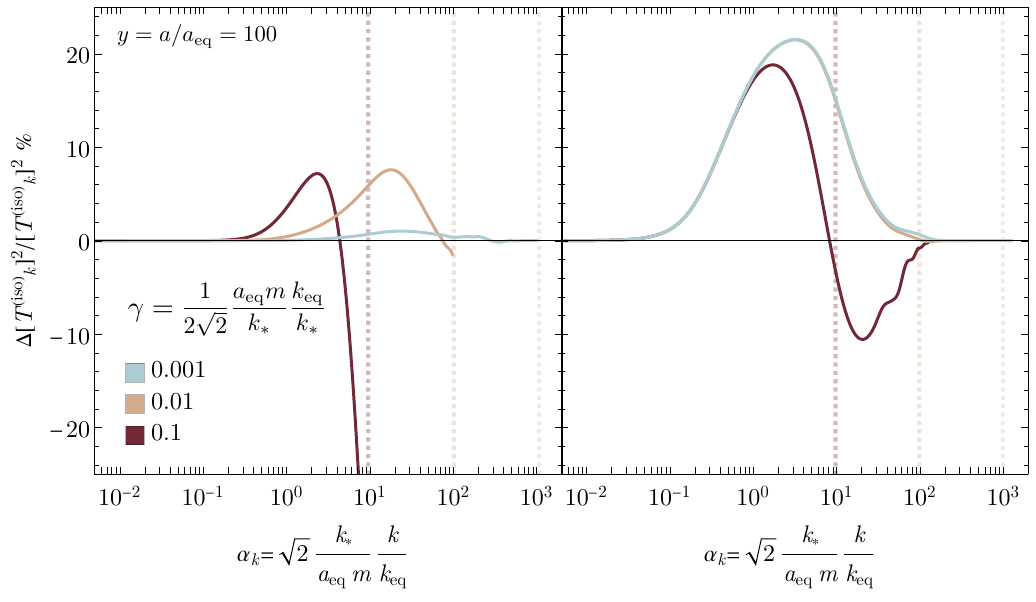}

    \caption{Each panel shows the fractional difference in the growth of isocurvature perturbations compared to the quasi-particle calculation of \cite{Amin:2025dtd}, for different choices of $f_0(p)$. Here, $[T^{(\mathrm{iso})}_k(y,y_0)]^2=P^{(\mathrm{iso})}_\delta(y,k)/P^{(\mathrm{iso})}_\delta(y_0,k)$, and $\gamma$ is a measure of wave effects. There are two types of deviations from the quasi-particle results: The first type appears at $\gamma \alpha_k \gtrsim 1$ (equivalently $k\gtrsim k_*$) due to the sines and cosines appearing in the free-streaming kernels. This effect vanishes as $\gamma \rightarrow 0$. The second occurs at $k \gtrsim k_{\mathrm{J}}^{\mathrm{eq}}$ due to $f_0^2$ instead of $f_0$ appearing in $T^{\mathrm{fs}\,(\mathrm{c})}$; this effect \emph{typically} survives when $\gamma \rightarrow 0$. \textit{Left panel:} For $f_0(p) \propto \Theta(k_*-p)$, \emph{both} deviations vanish as $\gamma \rightarrow 0$. The vertical dashed lines indicate $k=2k_*$, where the initial density power spectrum vanishes. For the \textit{right panel}, we used $f_0(p) \propto e^{-p^2/2k_*^2}$. Here, the $f_0^2$ effect persists for $\gamma \rightarrow 0$ (blue curve). The oscillations at $\gamma \alpha_k \gtrsim 1$ are due to the sines and cosines; their amplitude vanishes as $\gamma \rightarrow 0$. Expressions for initial power spectra and free-streaming kernels for these examples are provided in \cref{sec:examples}.}
    \label{fig:plotPSIsoCompareParticleVsWave}
\end{figure}

The second difference with the quasi-particle picture is the fact that initial Poisson fluctuations originate from wave interference, rather than the small number density of quasi-particles. Two waves with $p \sim k_*$ must be superposed to create a long-wavelength density perturbation. This manifests itself in the appearance of two factors of $f_0$ in $T^{{\mathrm{fs}}\,(\mathrm{c})}_k$, which is replaced by $T^{\mathrm{fs}\,(\mathrm{a})}_k$ for quasi-particles. Hence, depending on the shape of $f_0(p)$ the way these seed perturbations free-stream differs from quasi-particles. For instance, if $f_0$ is Gaussian the momenta in the superposition of waves dominating the Poisson noise are lower than if $f_0$ were a uniform sphere with the same velocity dispersion.

There is also the obvious deviation from the quasi-particle limit in that the field and density power spectra have a cutoff around $k \sim k_*$. \Cref{fig:plotPSIsoCompareParticleVsWave} shows the former two effects on the growth of the isocurvature spectrum. A similar analysis can be done for the free-streaming suppression of the adiabatic part.

\section{Numerical Simulations}
\label{sec:Numerics}
In this section we describe the algorithms to set up both initial conditions and evolution of the Schrödinger field. Further details can be found in \cref{sec:NumericalDetails}.

\subsection{Initial Conditions}
The initial conditions for warm wave dark matter including adiabatic perturbations is different from existing literature (which has focused on ``cold" wave dark matter with a dominant zero mode). The initial condition setup we use here is a generalization of \cite{Ling:2024qfv} to include bulk velocity perturbations, and is adapted for Schrödinger–Poisson as opposed to Klein–Gordon–Einstein simulations.

We start by generating a realization of the  density ($\delta_{\mathrm{ad}}(\bx)$) and velocity fields ($\bv_{\mathrm{ad}}(\bx)$) on large, but sub-horizon length scales based on the curvature spectrum from inflation after horizon entry. We then generate a field configuration $\psi(\bx)$ from any power spectrum $P_\psi(q)$ peaked at $k_*\gg aH$, such that the variance of the field is position-dependent with the position dependence determined by $\delta_{\mathrm{ad}}(\bx)$. The complex field $\psi(\bx)$ thus obtained is then rotated using $\bv_{\mathrm{ad}}(\bx)$. This procedure results in the generated field configurations having the desired $\delta_{\mathrm{ad}}(\bx)$ and $\bv_{\mathrm{ad}}(\bx)$ on large length scales (after ensemble averaging over different field realizations). On small length scales, the effects of $\delta_{\mathrm{ad}}$ and $\bv_{\mathrm{ad}}$ are negligible, and we naturally get an isocurvature density spectrum peaked at roughly the same scale as $P_\psi(q)$.
This initial condition generation is provided as a step-by-step algorithm in \cref{sec:AlgoIC}.

\subsection{Numerical Evolution}
\label{sec:numerics-evolution}

\begin{figure}
    \vspace*{-9ex}

    \centering
    \raisebox{-0.5\height}{\includegraphics{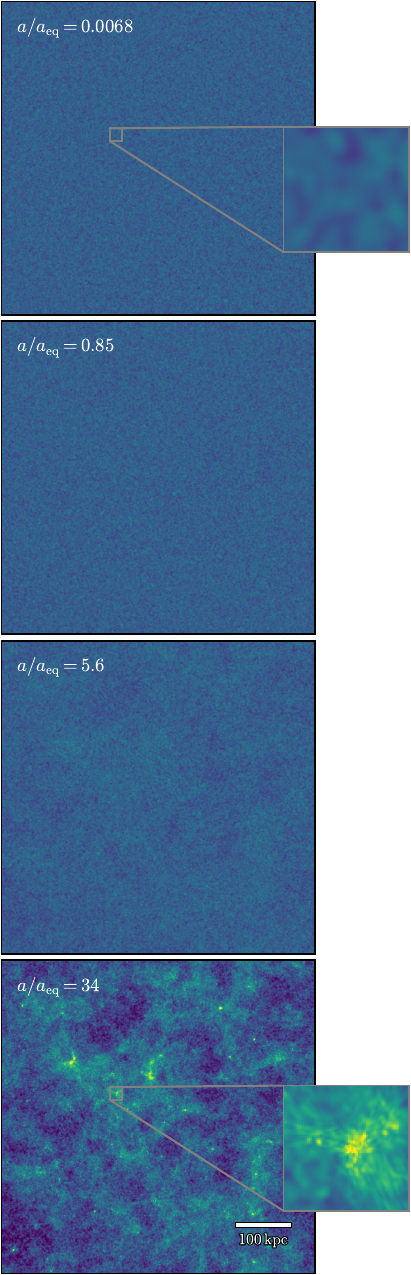}}%
    \hfill%
    \raisebox{-0.5\height}{\includegraphics{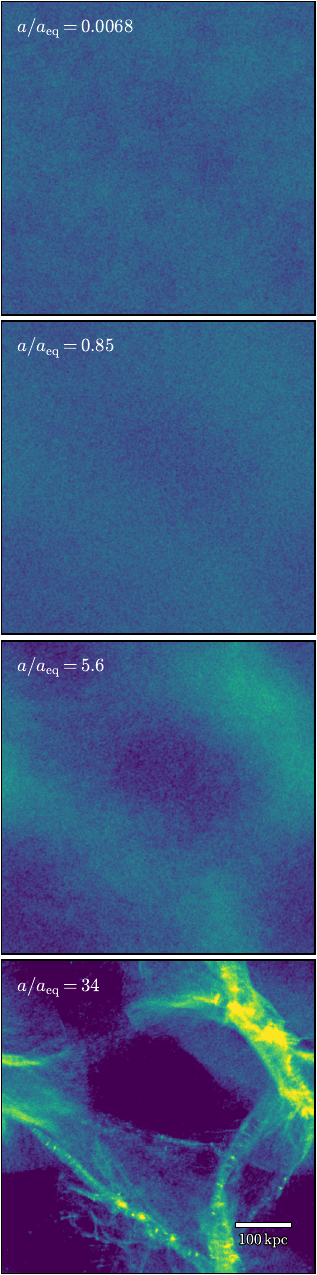}}%
    \,%
    \raisebox{-0.5\height}{\includegraphics{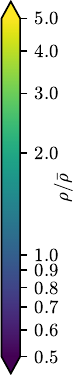}}

    \caption{%
        Evolution of the projected density field in simulations of warm wave dark matter.
        \textit{Left column:} Isocurvature ICs only. The de Broglie-scale variations in density are visible in the ICs. The formation of halos and solitons during matter domination are visible at late times.
        \textit{Right column:} ICs including both isocurvature and enhanced adiabatic perturbations. The erasure of some of the adiabatic perturbations is visible at early times, followed by growth of structure during matter domination.
    }
    \label{fig:sim-projections}
\end{figure}

We use the \code{AxiREPO} code \cite{May:2021wwp, May:2022gus} to perform the numerical simulations, evolving the Schrödinger–\allowbreak Poisson system of equations, \cref{eq:SPscalar}, across cosmic time in three spatial dimensions. \code{AxiREPO} implements a spectral method for the spatial derivatives and employs the ``kick-drift-kick" leapfrog algorithm for the time integration. A detailed overview of the numerical evolution algorithm and resolution-related considerations relevant for this work are discussed further \cref{sec:AxiREPO}. The initial conditions were generated as described in \cref{sec:AlgoIC}.

Cosmological parameters are taken from \cite{Planck:2015fie}, i.\,e.\ $\Omega_\Lambda = \num{0.6911}$, $\Omega_{\mathrm{m}} = \num{0.3089}$, $h = \num{0.6774}$.
For simplicity, we do not explicitly include baryons in our simulations, so $\Omega_{\mathrm{b}} = 0$.
We run the simulation from $a_0 / a_{\mathrm{eq}} = 0.006766$ (corresponding to an initial redshift of $z_0 = \num{5e5} - 1$) to $a_{\mathrm{f}} / a_{\mathrm{eq}} = 33.7$ (a final redshift of $z_{\mathrm{f}} = 99$). We note that resolution effects can set in at slightly earlier times than $z_{\mathrm{f}}$, especially in lower-resolution cases.

For a correct treatment of the evolution before matter–radiation equality, we included the impact of radiation ($\Omega_{\mathrm{r}}$) on the evolution of the scale factor via the Hubble expansion rate, \cref{eq:Hubble}.
For initial conditions with an adiabatic component, we artificially boost the value of the adiabatic power spectrum (following \cite{Ling:2024qfv, Amin:2025dtd}) such that $(k^3/2\pi^2)P^{(\mathrm{ad})}_\delta(k)|_{k = \SI{1}{\per\Mpc}} = \num{e-4}$.
This is done out of necessity due to computational limitations on the box size.
Without this boost, the contribution of the part of the adiabatic power spectrum accessible within the simulation box would be negligible compared to the isocurvature part, and we would not be able to study the behavior of both components in combination.

See \cref{tab:simulations} in \cref{sec:NumericalDetails} for a summary of performed simulations. We use either pure isocurvature or combined adiabatic and isocurvature initial conditions.
Simulations were run at different resolutions and box sizes in order to check numerical convergence.

\subsection{Simulation Results}
\label{sec:SimResults}

In this subsection we provide a comparison between analytical and numerical results. We will compare density power spectra and halo mass functions, and discuss the formation of solitons.

\subsubsection{Projected Density Field}

Snapshots of the density field at different redshifts are shown in \cref{fig:sim-projections}. The results in the left column did not include adiabatic initial power. The initial variations of the density field are on the de Broglie scale $k_*^{-1}$. After matter–radiation equality we see scale-dependent growth of the density field. The growth is suppressed below the Jeans length $k_{\mathrm{J}}^{-1}$, above which standard isocurvature growth is present. We discuss this more quantitatively when comparing the numerical and analytical power spectra as well as the halo mass functions.

For the second column, we included enhanced adiabatic perturbations. Here, we see that adiabatic perturbations on length scales smaller than the free-streaming scale $k_{\mathrm{fs}}^{-1}$ are erased during radiation domination. At later times, nonlinear structure forms on large length scales comparable to the simulation volume. This is due to the artificially enhanced adiabatic power. Again, features related to free-streaming suppression of the adiabatic power, and growth of the isocurvature part will be more easily seen in the power spectra comparisons.

\begin{figure}
    \centering
    \includegraphics[width=0.75\linewidth]{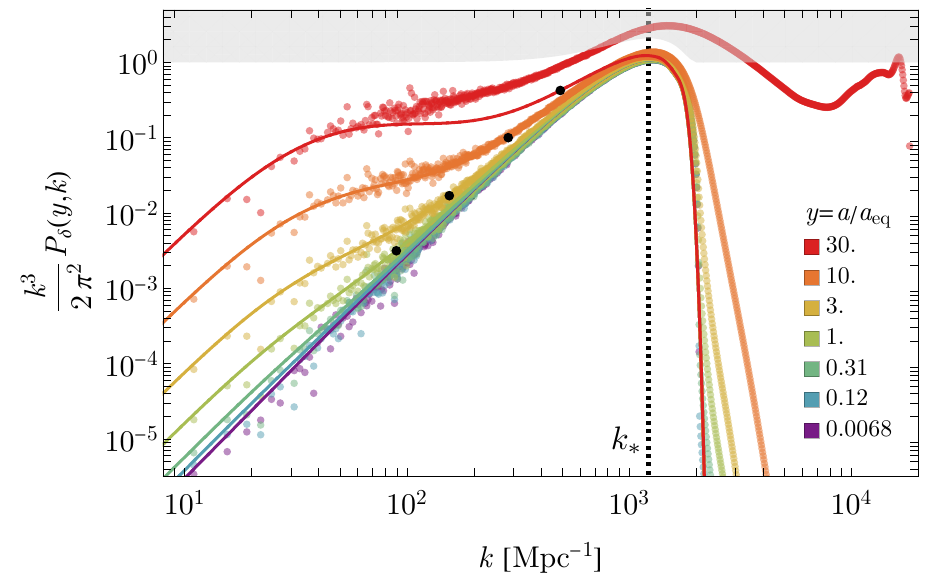}

    \caption{A comparison between the analytically predicted evolution of the matter power spectrum (solid lines) and results from Schrödinger–Poisson simulations (dots), \emph{without} including adiabatic initial conditions. The solid black circles on the analytic lines are the Jeans wavenumber $k_{\mathrm{J}}(y)$. The match with numerical results is excellent before the evolution becomes nonlinear at late times.}
    \label{fig:plotNumComparePSIso}
\end{figure}

\subsubsection{Power Spectra Comparison}
For the purpose of comparison, it is convenient to shift to $y=a/a_{\mathrm{eq}}$ as the time variable. In a universe with matter and radiation, the cosmic time $t$, $\eta$ and $y$ are connected via
\beq
    \label{eq:etaDef}
    \frac{\dl t}{a^2(t)} = \dl\eta =\frac{\sqrt{2}}{a_{\mathrm{eq}}k_{\mathrm{eq}}}\frac{\dl y}{y\sqrt{1+y}},
    \quad \text{with} \quad
    y=\frac{a}{a_{\mathrm{eq}}},
\eeq
with the Hubble expansion rate given by
\beq
    \label{eq:Hubble}
    H(y)=\frac{k_{\mathrm{eq}}}{\sqrt{2}a_{\mathrm{eq}}}y^{-2}\sqrt{1+y}\,.
\eeq
The scale factor at matter–radiation equality is $a_{\mathrm{eq}}\approx 1/3383$, and the comoving wavenumber associated with the horizon size at that time is $k_{\mathrm{eq}} = a_{\mathrm{eq}}H(a_{\mathrm{eq}})\approx \SI{0.01}{\per\Mpc}$ \cite{Planck:2018vyg}. In terms of these parameters, our Jeans and free-streaming wavenumbers (see \cref{eq:kfs,eq:kJ}) can be written as
\beq
    k_{\mathrm{J}}(y)&=\frac{\sqrt{3y}}{2}\frac{k_{\mathrm{eq}}}{\sigma_{\mathrm{eq}}}\approx \SI{120}{\per\Mpc} \sqrt{y}\left(\frac{\SI{22}{\km\per\s}}{\sigma_{\mathrm{eq}}}\right),\\
    k_{\mathrm{fs}}(y)&=\frac{1}{\sqrt{2}\mathcal{F}(y,y_0)}\frac{k_{\mathrm{eq}}}{\sigma_{\mathrm{eq}}}\approx \SI{15}{\per\Mpc} \frac{\mathcal{F}(1, 10^{-3})}{\mathcal{F}(y,y_0)}\left(\frac{\SI{22}{\km\per\s}}{\sigma_{\mathrm{eq}}}\right),
\eeq
with $\sigma_{\mathrm{eq}}=\sigma(\eta_{\mathrm{eq}})$ and $\mathcal{F}(y,y_0) = \ln\ml[(y/y_0)(1+\sqrt{1+y_0})^2 / (1+\sqrt{1+y})^2\mr]$.\footnote{The free-streaming scale would change if we moved away from the standard expansion history, see \cite{Long:2024imw}.}

\begin{figure}
    \centering
    \includegraphics[width=0.95\linewidth]{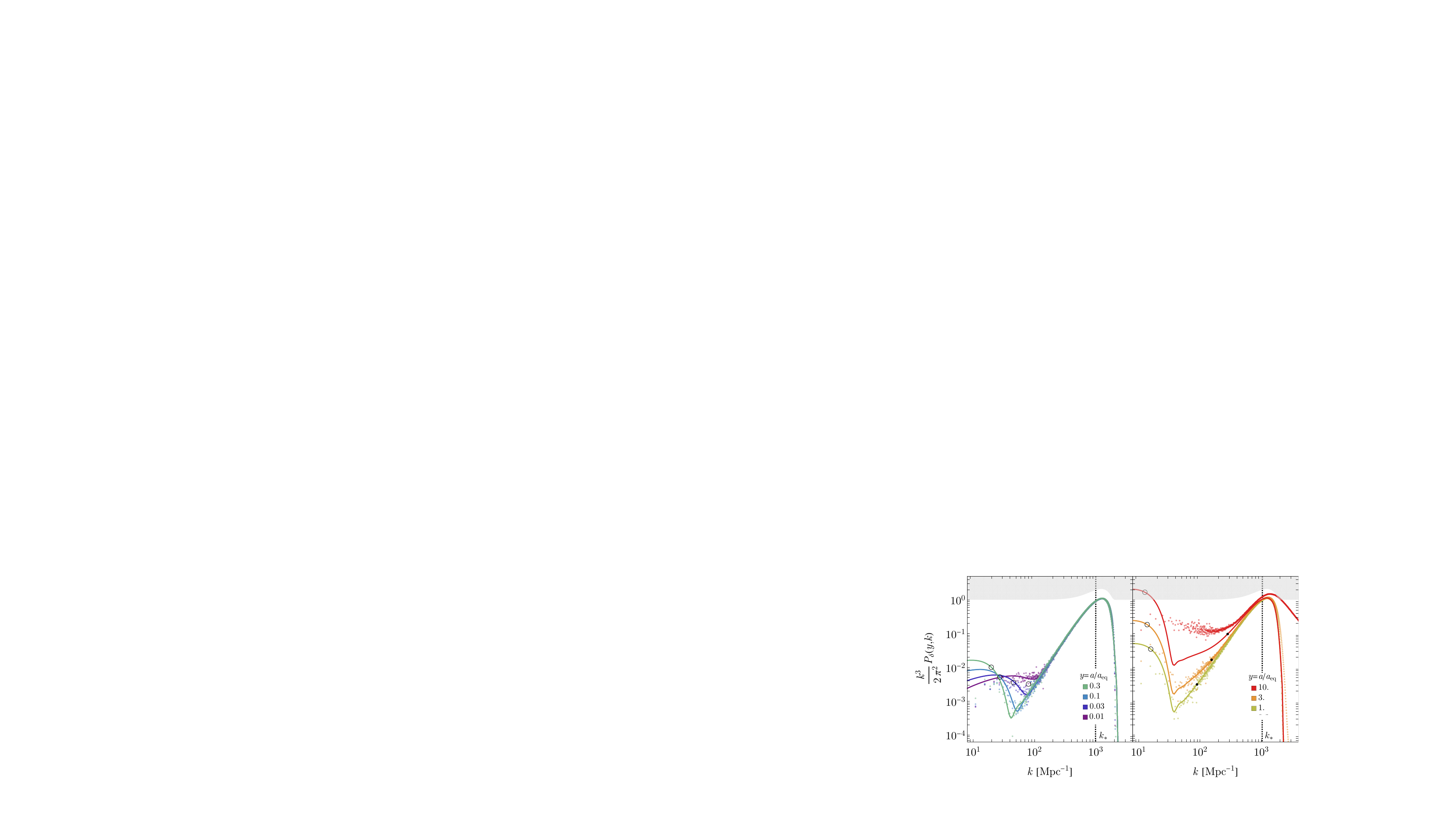}

    \caption{A comparison between the analytically predicted evolution of the matter power spectrum (solid lines) and results from Schrödinger–Poisson simulations (dots), \emph{including} enhanced adiabatic initial conditions. The solid black circles on the analytic lines are the Jeans wavenumber $k_{\mathrm{J}}(y)$, whereas the open circles are the free-streaming wavenumber $k_{\mathrm{fs}}(y)$. The match with numerical results is very good below $k_*$ and before the density perturbation evolution becomes nonlinear (gray region). On the left, note that free-streaming erases adiabatic perturbations, leaving the uncorrelated white-noise like part unchanged during radiation domination. During matter domination (right), all scales below the Jeans scales grow. The artificially enhanced adiabatic initial conditions lead to large length-scales becoming nonlinear first.}
    \label{fig:plotNumComparePSAdIso}
    \end{figure}

In \cref{fig:plotNumComparePSIso,fig:plotNumComparePSAdIso}, we show a comparison between the power spectra obtained from Schrödinger–Poisson simulations, and our analytical results for the same. The agreement is excellent before the density perturbations become nonlinear.

\subsubsection{Halo Mass Function}
We use the linear power spectrum to calculate the mass function of collapsed halos in the nonlinear regime. The differential fraction of dark matter in halos of mass $M$ (per logarithmic mass interval) can be estimated via
\beq
    \label{eq:dfdlnM}
    \frac{\dl f}{\dl\ln(M)} = 0.658\sigma_M^{-0.582}e^{-1.056/\sigma_M^2} \ml|\frac{\dl\ln(\sigma_M)}{\dl\ln(M)}\mr|,
    \quad
    \sigma^2_M = \int_0^{k_\mathrm{max}} \! \dl\ln(k) \frac{k^3}{2\pi^2} P_\delta(k) W^2(k R_M).
\eeq
The fitting formula for $\dl f/\dl\ln(M)$ is based on excursion set theory calculations \cite{Delos:2023eve}, $\sigma_M^2$ is the root mean square density variance in spheres of mass $M$, $R_M\equiv[{3M}/{(4\pi\bar{\rho})}]^{1/3}$ is the radius of such a sphere and $W(z) \equiv 3\left(\sin(z) - z\cos(z)\right)/z^3$ is the spherical top-hat window function in Fourier space. We take the upper limit $k_{\mathrm{max}} = k_{\mathrm{J}}(\eta)/4$ reflecting the expected suppression of halo formation due to the Jeans scale. See \cite{Amin:2025dtd} for a more detailed discussion.

\begin{figure}
    \centering
    \includegraphics[width=0.8\linewidth]{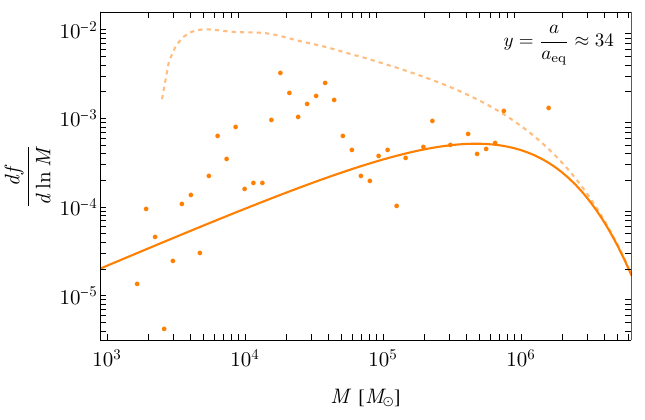}

    \caption{Comparison between the differential halo mass function obtained using our analytically predicted power spectrum (solid orange) with results from an Schrödinger–Poisson simulations (dots). These are for the case of isocurvature-only initial conditions.
    Solid lines are our analytic predictions with power spectrum cut off above $k_{\mathrm{J}}/4$. The lighter dashed lines show analytic predictions when this Jeans scale related cutoff is not imposed. The halos typically contain a soliton at the center, although the the solitons are not spatially resolved at the high-mass end.}
    \label{fig:plotMassFuncCompare}
\end{figure}

In \cref{fig:plotMassFuncCompare}, the mass function predicted by \cref{eq:dfdlnM} matches the simulation results well at the high-mass end. The results for the halo mass function when $k_{\mathrm{max}} \rightarrow \infty$ is shown as a dashed curve, which does not fit the simulations results as well.

For the simulation data points, the algorithm described in \cite{May:2021wwp,May:2022gus} was used to identify collapsed objects (halos). We first find all points with $\rho/\bar{\rho} \ge 60$. Adjacent points satisfying this criterion are identified as being part of the same object. Then, starting with the highest-density point, we identify the sphere whose average density is $200$ times the background density $\bar{\rho}$ at that time. The mass within this region is $M_{200}$.

There is deviation from the predicted mass function in the low- to mid-range of $M_{200}$. This could arise due to formation of solitons which may not be well captured by excursion set theory/Press–Schechter-type arguments.

\subsubsection{Halos and Solitons}
Nonlinear dynamics will become relevant when
\beq
    \label{eq:NLcond}
    \frac{k^3}{2\pi^2} \left[P_{\delta}(y,k) - P_{\mathrm{iso}}(y_0,k)\right]_{y=y_{\mathrm{nl}}} \sim 1 .
\eeq
If $\gamma\sim k_{\mathrm{J}}^{\mathrm{eq}}/k_*\ll 1$, then using \cref{eq:NLcond,eq:PisoScalings} the scale $k_{\mathrm{J}}^{\mathrm{eq}}$ will become nonlinear first at time
\beq
    \label{eq:ynlysol}
    y_{\mathrm{nl}} \sim
    \left(\frac{k_*}{k_{\mathrm{J}}^{\mathrm{eq}}}\right)^{3/2} \sim
    \gamma^{-3/2} .
\eeq
This will lead to the formation of virialized halos with mass $M \sim \bar{\rho} (a_{\mathrm{nl}}/k_{\mathrm{J}}^{\mathrm{eq}})^3$. On the other hand, linear theory predicts soliton formation when the fuzzy Jeans scale $k_{\mathrm{j}}(y)\sim \sqrt{k_{\mathrm{J}}(y) k_*}$ becomes nonlinear. This happens when
\beq
    \label{eq:solcond}
    k_{\mathrm{J}}(y)|_{y=\bar{y}_{\mathrm{sol}}} \sim k_* \,,
    \quad \text{so that} \quad
    \bar{y}_{\mathrm{sol}} \sim {\gamma}^{-2},
\eeq
with soliton mass $M\sim \bar\rho (a_{\mathrm{sol}}/k_*)^3$. For $\gamma \ll 1$, we have $y_{\mathrm{nl}} \ll \bar{y}_{\mathrm{sol}}$. So in that case halos form first, and the evolution of shorter scales is expected to significantly deviate from our linear analysis. In contrast, when $\gamma \sim 1$ as in \cite{Gorghetto:2022sue, Chang:2024fol, Gorghetto:2024vnp} a significant fraction of mass ends up in solitons because $y_{\mathrm{nl}} \sim \bar{y}_{\mathrm{sol}}$.

For our fiducial parameters, $\gamma \approx 0.02$. This is small, but not small enough for the asymptotic formula, \cref{eq:PisoScalings}, to be rigorous, and to have a big hierarchy between the two time scales. From our simulations, we see evidence of nonlinear dynamics at $y_{\mathrm{nl}} \approx 20$. At this time, we see tentative evidence for $k\sim k_{\mathrm{J}}$ becoming nonlinear, along with a growth of the peak at $k\sim k_*$. These together are seen to lead to the formation of halos on length scales $2\pi a_{\mathrm{nl}}/k_{\mathrm{J}}^{\mathrm{eq}}$ with solitons of size $\sim 2\pi a_{\mathrm{nl}}/k_*$ inside them.

\begin{figure}
    \centering
    \raisebox{-0.5\height}{\includegraphics[width=0.5\linewidth]{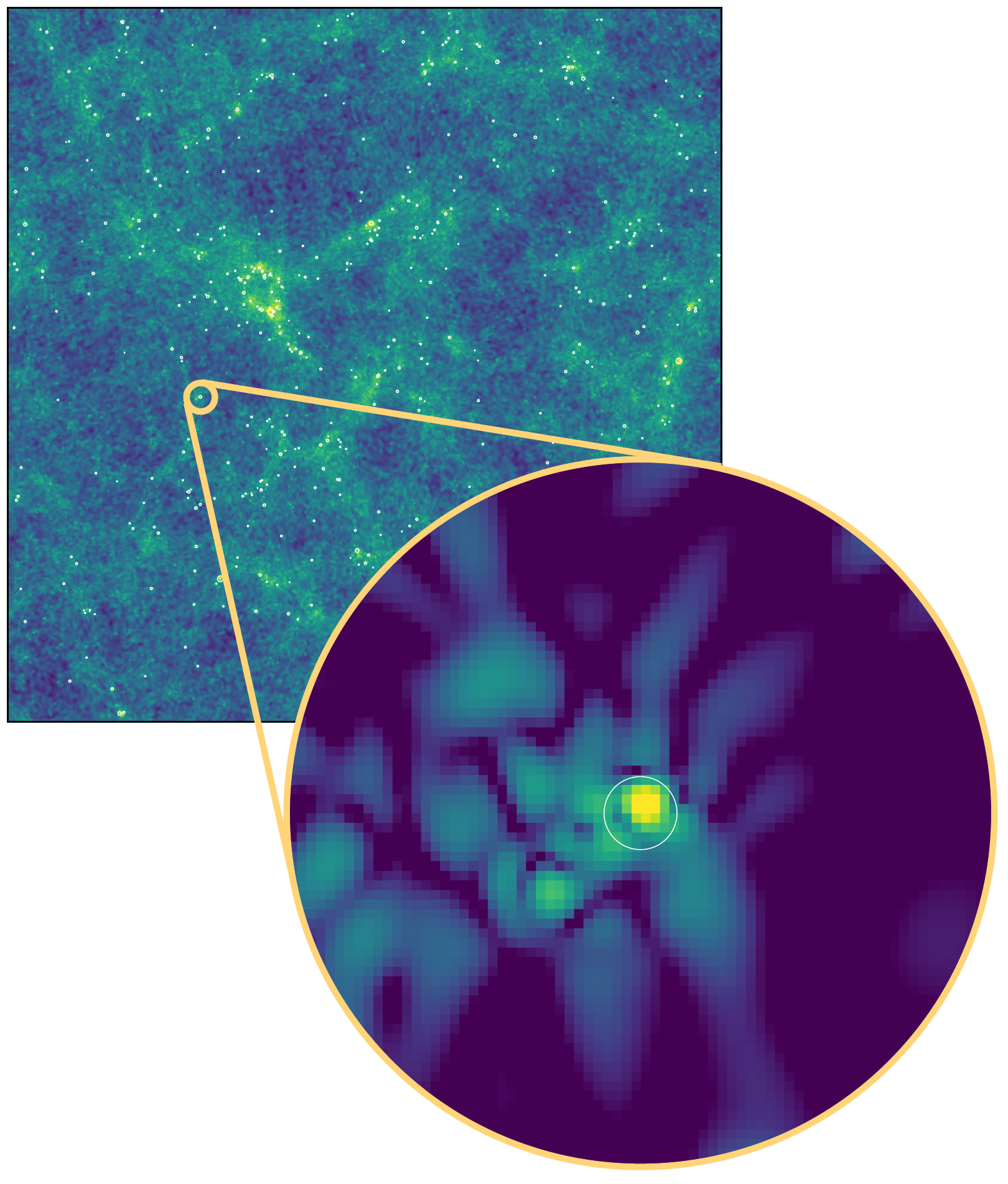}}%
    \hfill%
    \raisebox{-0.5\height}{\includegraphics{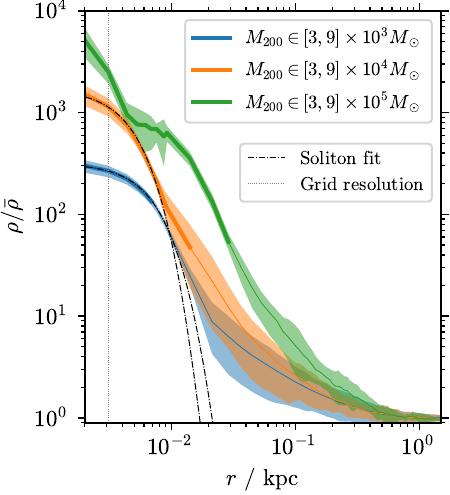}}

    \caption{\textit{Left panel:} A density projection through \SI{30}{\percent} of the simulation volume (square), and a zoomed-in density slice showing a soliton and surrounding halo (circle inset). Identified objects are marked with white dots/circles. The \textit{right panel} shows extracted soliton and halo profiles (after angular averaging) from the simulations (solid colored lines) with the shaded bands around solid lines containing $\sim \SI{85}{\percent}$ of sample profiles. The dashed lines show the soliton profiles. Low-mass solitons (blue, orange) are spatially resolved, whereas higher mass ones (green) are not. At the final time in our simulations, the estimated mass in all solitons is comparable to the sum of $M_{200}$ of all objects.}
    \label{fig:SolitonProfiles}
\end{figure}

We investigated density profiles of collapsed objects at $y_{\mathrm{f}} \approx 34$, and most have soliton-like configurations in their centers. The lightest identified objects in our simulations (with $M_{200} \sim \SI{e3}{\Msun}$, \cref{fig:plotMassFuncCompare}) have a soliton in the center with a physical radius $R_{\mathrm{sol}} \sim \SI{0.01}{\kpc}$, surrounded by interference granules. Using $R_{\mathrm{sol}} \approx 10 (\hbar/mc)^2/(GM_{\mathrm{sol}}/c^2)$, we have $M_{\mathrm{sol}} \sim \SI{e3}{\Msun}$ which is the same order of magnitude as the host halo. These light solitons are well-resolved in our simulations.\footnote{Note that our (physical) spatial resolution at $y_{\mathrm{f}}$ is $a_{\mathrm{f}} \Delta x \approx \SI{0.003}{\kpc}$.} As we move to higher $M_{200}$ (closer to \SIrange{e5}{e6}{\Msun}), we do not have sufficient resolution to resolve the (potential) solitons at the center, although the density profiles show a clear deviation from the purely Navarro–Frenk–White (NFW) shape expected in the absence of wave effects. Some sample profiles are shown in \cref{fig:SolitonProfiles}.

By identifying the central density of collapsed regions at the final time in the simulations as central soliton densities, we found that the total mass in solitons is about \SI{50}{\percent} of the mass in collapsed objects ($M_{200}$). The soliton fraction of total mass in the simulation volume (not just in $M_{200}$) at this time is much smaller, $\sim \SI{0.1}{\percent}$. While the mass fraction in halos is similarly small at this early time, it is expected that it will saturate to order unity, see e.\,g.\ \cite{Gorghetto:2024vnp}. Determining the final mass function of solitons is an interesting question that we leave for future work due to numerical limitations.

\section{Discussion \& Summary}
\label{sec:summary}
When light dark matter is produced via causal processes after inflation, the dark matter density can be dominated by sub-Hubble field modes. These modes lead to an isocurvature enhancement of the power spectrum at small scales and a free-streaming suppression in the adiabatic part of the spectrum compared to the standard CDM scenario. In this work, we have explored the growth of structures in these scenarios in the radiation- and matter-dominated eras. We provided an analytical understanding of the scale-dependent growth of the density power spectrum in the linear regime, and an approximate understanding of the halo mass function in the nonlinear one. We carried out new cosmological simulations of the Schrödinger–Poisson system including adiabatic and isocurvature initial conditions relevant for our scenario of interest. We also explored the formation of solitons and halos resulting from the evolution of the isocurvature part of the spectrum.

Our present work finds its motivation in \cite{Amin:2022nlh}. It relies on and builds upon our recent earlier works \cite{Amin:2025dtd} and \cite{Ling:2024qfv}. In particular, compared to \cite{Amin:2025dtd}, we included additional wave-dynamical effects in the analytic derivation, and also carried out wave-dynamical simulations.

We summarize our key results below:
\begin{enumerate}
    \item \textbf{Analytical Results:} Starting with the Schrödinger–Poisson system, we provided a derivation of the scale-dependent evolution of the dark matter density power spectrum. The input for the derivation is an initial power spectrum of the field which is peaked at $k \sim k_*$. The evolution requires solving a Volterra-type equation (see \cite{Amin:2025dtd} for an efficient algorithm). We found the following:
    \begin{enumerate}
        \item \textbf{Isocurvature:} The isocurvature part of the spectrum does not evolve significantly during radiation domination. During matter domination, the spectrum grows $\propto (a/a_{\mathrm{eq}})^2$ for $k < k_{\mathrm{J}}(a_{\mathrm{eq}})$, where $k_{\mathrm{J}}(a_{\mathrm{eq}}) \sim k_{\mathrm{eq}}/(k_*/(a_{\mathrm{eq}} m))$ is the Jeans scale at equality. For $k > k_{\mathrm{J}}(a_{\mathrm{eq}})$, the growth is scale-dependent with essentially no growth until $k < k_{\mathrm{J}}(a) \propto \sqrt{a/a_{\mathrm{eq}}} \, k_{\mathrm{J}}(a_{\mathrm{eq}})$.
        \item \textbf{Adiabatic:} By matter–radiation equality, the adiabatic part of the density spectrum is erased for $k > k_{\mathrm{fs}}(a_{\mathrm{eq}}) \sim k_{\mathrm{J}}(a_{\mathrm{eq}})/\ln(m/H_{\mathrm{eq}})$, where $k_{\mathrm{fs}}$ is the free-streaming scale at equality. For $k < k_{\mathrm{fs}}(a)\sim k_{\mathrm{fs}}(a_{\mathrm{eq}})$ the evolution of the power spectrum during matter domination is again $\propto (a/a_{\mathrm{eq}})^2$ as in the CDM case.
        \item \textbf{Wave-dynamical effects beyond quasi-particles:} In the present work, wave effects which impact scales $k\sim k_*$, and are quantified by the ratio $\gamma\sim k_{\mathrm{J}}(a_{\mathrm{eq}})/k_*$, are also included. These effects go beyond the quasi-particle picture of \cite{Amin:2025dtd}, but vanish as $\gamma \rightarrow 0$ ($k_* \rightarrow \infty$ at fixed $k_*/m$). Additional wave interference-related deviations ($\sim \SI{10}{\percent}$) from the quasi-particle limit, which can impact scales $k\gtrsim k_{\mathrm{J}}(a_{\mathrm{eq}})$, can still remain even in the $\gamma\rightarrow 0$ limit.
    \end{enumerate}
    \item \textbf{Simulations \& Nonlinear Evolution:} We provided $3+1$-dimensional Schrödinger–\allowbreak Poisson simulations of warm wave dark matter, including sub-horizon isocurvature (with a peaked spectrum) and adiabatic initial conditions. These were cosmological simulations, spanning the radiation and matter-dominated eras, with comoving simulation volumes $L^3 \sim (\SI{0.5}{\Mpc})^3$ and numerical grid sizes $N^3 \sim 2000^3$. We found the following:
    \begin{enumerate}
        \item \textbf{Density Power Spectra:} The density power spectra extracted from simulations matched the analytical predictions in the linear regime. In particular, we saw the free-streaming suppression of the adiabatic part and the effects of the Jeans scale on the isocurvature growth during matter domination.
        \item \textbf{Halo Mass Function:} The halo mass functions extracted from simulations were roughly consistent with our ansatz based on an analytically computed linear matter power spectrum, at least at the high-mass end.
        \item \textbf{Halos and Solitons:} Typically, the isocurvature part of the spectrum becomes nonlinear first at $a/a_{\mathrm{eq}} \sim {\gamma}^{-3/2}$, where ${\gamma}\sim k_{\mathrm{J}}(a_{\mathrm{eq}})/k_*$. This leads to the formation of halos (above the Jeans length) with solitons at their centers. At the end of our (isocurvature-only) simulations, $a/a_{\mathrm{eq}} \approx 34$, we estimate that solitons formed an order-unity fraction of the total mass in collapsed halos (but only $\sim \SI{0.1}{\percent}$ of the total mass in the simulation volume).
    \end{enumerate}
\end{enumerate}

For our formalism and simulations, we had ultralight, warm wave dark matter in mind because of the observationally accessible features on quasi-linear scales. The formalism also applies to cases of wave dark matter where either the white noise (with or without Jeans-related effects) or free streaming are not necessarily relevant on the same scales (e.\,g.~\cite{McQuinn, Gorghetto:2020qws}).

Using the results in the present paper, and the efficient evaluation of power spectra \cite{Amin:2025dtd}, we can reduce theoretical uncertainties in obtaining observational constraints on the warm wave dark matter mass and production mechanism. Beyond the power spectrum, in an upcoming paper we will provide detailed predictions of halo and soliton formation times and distributions (along with effects on and from baryons) which could provide exciting avenues for observational probes in the nonlinear regime \cite{Drlica-Wagner:2022lbd}. 

\section{Acknowledgements}
We thank Christian Capanelli, Wayne Hu, Fabian Schmidt and Huangyu Xiao for helpful conversations. We thank Siyang Ling for early discussions of the initial conditions setup, and Sten Delos for providing a cross-check for the power spectra evolution via a modification of the algorithm in \cite{Amin:2025dtd} to include wave dynamical effects. MAA is supported by DOE grant DE-SC0021619.
The numerical simulations were performed on systems at the Max Planck Computing and Data Facility.

\bibliographystyle{JHEP}
\bibliography{main}

\providecommand{\href}[2]{#2}\begingroup\raggedright\begin{thebibliography}{10}

\bibitem{Cirelli:2024ssz}
M.~Cirelli, A.~Strumia and J.~Zupan, \emph{{Dark Matter}},  \href{https://arxiv.org/abs/2406.01705}{{\ttfamily 2406.01705}}.

\bibitem{Dalal}
N.~Dalal and A.~Kravtsov, \emph{{Not so fuzzy: excluding FDM with sizes and stellar kinematics of ultra-faint dwarf galaxies}},  \href{https://arxiv.org/abs/2203.05750}{{\ttfamily 2203.05750}}.

\bibitem{Amin:2022nlh}
M.A.~Amin and M.~Mirbabayi, \emph{{A Lower Bound on Dark Matter Mass}}, \href{https://doi.org/10.1103/PhysRevLett.132.221004}{\emph{Phys. Rev. Lett.} {\bfseries 132} (2024) 221004} [\href{https://arxiv.org/abs/2211.09775}{{\ttfamily 2211.09775}}].

\bibitem{ParticleDataGroup:2020ssz}
{\scshape Particle Data Group} collaboration, \emph{{Review of Particle Physics}}, \href{https://doi.org/10.1093/ptep/ptaa104}{\emph{PTEP} {\bfseries 2020} (2020) 083C01}.

\bibitem{Marsh:2015xka}
D.J.E.~Marsh, \emph{{Axion Cosmology}}, \href{https://doi.org/10.1016/j.physrep.2016.06.005}{\emph{Phys. Rept.} {\bfseries 643} (2016) 1} [\href{https://arxiv.org/abs/1510.07633}{{\ttfamily 1510.07633}}].

\bibitem{Hui:2016ltb}
L.~Hui, J.P.~Ostriker, S.~Tremaine and E.~Witten, \emph{{Ultralight scalars as cosmological dark matter}}, \href{https://doi.org/10.1103/PhysRevD.95.043541}{\emph{Phys. Rev. D} {\bfseries 95} (2017) 043541} [\href{https://arxiv.org/abs/1610.08297}{{\ttfamily 1610.08297}}].

\bibitem{Ferreira:2020fam}
E.G.M.~Ferreira, \emph{{Ultra-light dark matter}}, \href{https://doi.org/10.1007/s00159-021-00135-6}{\emph{Astron. Astrophys. Rev.} {\bfseries 29} (2021) 7} [\href{https://arxiv.org/abs/2005.03254}{{\ttfamily 2005.03254}}].

\bibitem{Preskill:1982cy}
J.~Preskill, M.B.~Wise and F.~Wilczek, \emph{{Cosmology of the Invisible Axion}}, \href{https://doi.org/10.1016/0370-2693(83)90637-8}{\emph{Phys. Lett. B} {\bfseries 120} (1983) 127}.

\bibitem{OHare:2024nmr}
C.A.J.~O'Hare, \emph{{Cosmology of axion dark matter}}, \href{https://doi.org/10.22323/1.454.0040}{\emph{PoS} {\bfseries COSMICWISPers} (2024) 040} [\href{https://arxiv.org/abs/2403.17697}{{\ttfamily 2403.17697}}].

\bibitem{Antypas:2022asj}
D.~Antypas et~al., \emph{{New Horizons: Scalar and Vector Ultralight Dark Matter}},  \href{https://arxiv.org/abs/2203.14915}{{\ttfamily 2203.14915}}.

\bibitem{Marzola:2017lbt}
L.~Marzola, M.~Raidal and F.R.~Urban, \emph{{Oscillating Spin-2 Dark Matter}}, \href{https://doi.org/10.1103/PhysRevD.97.024010}{\emph{Phys. Rev. D} {\bfseries 97} (2018) 024010} [\href{https://arxiv.org/abs/1708.04253}{{\ttfamily 1708.04253}}].

\bibitem{Jain:2021pnk}
M.~Jain and M.A.~Amin, \emph{{Polarized solitons in higher-spin wave dark matter}}, \href{https://doi.org/10.1103/PhysRevD.105.056019}{\emph{Phys. Rev. D} {\bfseries 105} (2022) 056019} [\href{https://arxiv.org/abs/2109.04892}{{\ttfamily 2109.04892}}].

\bibitem{Buschmann:2021sdq}
M.~Buschmann, J.W.~Foster, A.~Hook, A.~Peterson, D.E.~Willcox, W.~Zhang et~al., \emph{{Dark matter from axion strings with adaptive mesh refinement}}, \href{https://doi.org/10.1038/s41467-022-28669-y}{\emph{Nature Commun.} {\bfseries 13} (2022) 1049} [\href{https://arxiv.org/abs/2108.05368}{{\ttfamily 2108.05368}}].

\bibitem{Gorghetto}
M.~Gorghetto, E.~Hardy and G.~Villadoro, \emph{{More axions from strings}}, \href{https://doi.org/10.21468/SciPostPhys.10.2.050}{\emph{SciPost Phys.} {\bfseries 10} (2021) 050} [\href{https://arxiv.org/abs/2007.04990}{{\ttfamily 2007.04990}}].

\bibitem{Saikawa:2024bta}
K.~Saikawa, J.~Redondo, A.~Vaquero and M.~Kaltschmidt, \emph{{Spectrum of global string networks and the axion dark matter mass}}, \href{https://doi.org/10.1088/1475-7516/2024/10/043}{\emph{JCAP} {\bfseries 10} (2024) 043} [\href{https://arxiv.org/abs/2401.17253}{{\ttfamily 2401.17253}}].

\bibitem{Graham}
P.W.~Graham, J.~Mardon and S.~Rajendran, \emph{{Vector Dark Matter from Inflationary Fluctuations}}, \href{https://doi.org/10.1103/PhysRevD.93.103520}{\emph{Phys. Rev. D} {\bfseries 93} (2016) 103520} [\href{https://arxiv.org/abs/1504.02102}{{\ttfamily 1504.02102}}].

\bibitem{Kolb:2020fwh}
E.W.~Kolb and A.J.~Long, \emph{{Completely dark photons from gravitational particle production during the inflationary era}}, \href{https://doi.org/10.1007/JHEP03(2021)283}{\emph{JHEP} {\bfseries 03} (2021) 283} [\href{https://arxiv.org/abs/2009.03828}{{\ttfamily 2009.03828}}].

\bibitem{Agrawal}
P.~Agrawal, N.~Kitajima, M.~Reece, T.~Sekiguchi and F.~Takahashi, \emph{{Relic Abundance of Dark Photon Dark Matter}}, \href{https://doi.org/10.1016/j.physletb.2019.135136}{\emph{Phys. Lett. B} {\bfseries 801} (2020) 135136} [\href{https://arxiv.org/abs/1810.07188}{{\ttfamily 1810.07188}}].

\bibitem{Dror:2018pdh}
J.A.~Dror, K.~Harigaya and V.~Narayan, \emph{{Parametric Resonance Production of Ultralight Vector Dark Matter}}, \href{https://doi.org/10.1103/PhysRevD.99.035036}{\emph{Phys. Rev. D} {\bfseries 99} (2019) 035036} [\href{https://arxiv.org/abs/1810.07195}{{\ttfamily 1810.07195}}].

\bibitem{Long:2019lwl}
A.J.~Long and L.-T.~Wang, \emph{{Dark Photon Dark Matter from a Network of Cosmic Strings}}, \href{https://doi.org/10.1103/PhysRevD.99.063529}{\emph{Phys. Rev. D} {\bfseries 99} (2019) 063529} [\href{https://arxiv.org/abs/1901.03312}{{\ttfamily 1901.03312}}].

\bibitem{Adshead:2023qiw}
P.~Adshead, K.D.~Lozanov and Z.J.~Weiner, \emph{{Dark photon dark matter from an oscillating dilaton}}, \href{https://doi.org/10.1103/PhysRevD.107.083519}{\emph{Phys. Rev. D} {\bfseries 107} (2023) 083519} [\href{https://arxiv.org/abs/2301.07718}{{\ttfamily 2301.07718}}].

\bibitem{Cyncynates:2023zwj}
D.~Cyncynates and Z.J.~Weiner, \emph{{Detectable, defect-free dark photon dark matter}},  \href{https://arxiv.org/abs/2310.18397}{{\ttfamily 2310.18397}}.

\bibitem{McQuinn}
V.~Ir\v{s}i\v{c}, H.~Xiao and M.~McQuinn, \emph{{Early structure formation constraints on the ultralight axion in the postinflation scenario}}, \href{https://doi.org/10.1103/PhysRevD.101.123518}{\emph{Phys. Rev. D} {\bfseries 101} (2020) 123518} [\href{https://arxiv.org/abs/1911.11150}{{\ttfamily 1911.11150}}].

\bibitem{Long:2024imw}
A.J.~Long and M.~Venegas, \emph{{Free streaming of warm wave dark matter in modified expansion histories}},  \href{https://arxiv.org/abs/2412.14322}{{\ttfamily 2412.14322}}.

\bibitem{Ling:2024qfv}
S.~Ling and M.A.~Amin, \emph{{Free streaming in warm wave dark matter}}, \href{https://doi.org/10.1088/1475-7516/2025/02/025}{\emph{JCAP} {\bfseries 02} (2025) 025} [\href{https://arxiv.org/abs/2408.05591}{{\ttfamily 2408.05591}}].

\bibitem{Amin:2025dtd}
M.A.~Amin, M.S.~Delos and M.~Mirbabayi, \emph{{Structure Formation with Warm White Noise: Effects of Finite Number Density and Velocity Dispersion in Particle and Wave Dark Matter}},  \href{https://arxiv.org/abs/2503.20881}{{\ttfamily 2503.20881}}.

\bibitem{Schive:2014dra}
H.-Y.~Schive, T.~Chiueh and T.~Broadhurst, \emph{{Cosmic Structure as the Quantum Interference of a Coherent Dark Wave}}, \href{https://doi.org/10.1038/nphys2996}{\emph{Nature Phys.} {\bfseries 10} (2014) 496} [\href{https://arxiv.org/abs/1406.6586}{{\ttfamily 1406.6586}}].

\bibitem{Drlica-Wagner:2022lbd}
A.~Drlica-Wagner et~al., \emph{{Report of the Topical Group on Cosmic Probes of Dark Matter for Snowmass 2021}},  \href{https://arxiv.org/abs/2209.08215}{{\ttfamily 2209.08215}}.

\bibitem{Mondino:2020rkn}
C.~Mondino, A.-M.~Taki, K.~Van~Tilburg and N.~Weiner, \emph{{First Results on Dark Matter Substructure from Astrometric Weak Lensing}}, \href{https://doi.org/10.1103/PhysRevLett.125.111101}{\emph{Phys. Rev. Lett.} {\bfseries 125} (2020) 111101} [\href{https://arxiv.org/abs/2002.01938}{{\ttfamily 2002.01938}}].

\bibitem{Sabti:2021unj}
N.~Sabti, J.B.~Mu\~noz and D.~Blas, \emph{{New Roads to the Small-scale Universe: Measurements of the Clustering of Matter with the High-redshift UV Galaxy Luminosity Function}}, \href{https://doi.org/10.3847/2041-8213/ac5e9c}{\emph{Astrophys. J. Lett.} {\bfseries 928} (2022) L20} [\href{https://arxiv.org/abs/2110.13161}{{\ttfamily 2110.13161}}].

\bibitem{Gilman:2021gkj}
D.~Gilman, A.~Benson, J.~Bovy, S.~Birrer, T.~Treu and A.~Nierenberg, \emph{{The primordial matter power spectrum on sub-galactic scales}}, \href{https://doi.org/10.1093/mnras/stac670}{\emph{Mon. Not. Roy. Astron. Soc.} {\bfseries 512} (2022) 3163} [\href{https://arxiv.org/abs/2112.03293}{{\ttfamily 2112.03293}}].

\bibitem{Delos:2021ouc}
M.S.~Delos and F.~Schmidt, \emph{{Stellar streams and dark substructure: the diffusion regime}}, \href{https://doi.org/10.1093/mnras/stac1022}{\emph{Mon. Not. Roy. Astron. Soc.} {\bfseries 513} (2022) 3682} [\href{https://arxiv.org/abs/2108.13420}{{\ttfamily 2108.13420}}].

\bibitem{Boylan-Kolchin:2022kae}
M.~Boylan-Kolchin, \emph{{Stress testing \ensuremath{\Lambda}CDM with high-redshift galaxy candidates}}, \href{https://doi.org/10.1038/s41550-023-01937-7}{\emph{Nature Astron.} {\bfseries 7} (2023) 731} [\href{https://arxiv.org/abs/2208.01611}{{\ttfamily 2208.01611}}].

\bibitem{Chung:2023syw}
D.J.H.~Chung, M.~M\"unchmeyer and S.C.~Tadepalli, \emph{{Search for isocurvature with large-scale structure: A forecast for Euclid and MegaMapper using EFTofLSS}}, \href{https://doi.org/10.1103/PhysRevD.108.103542}{\emph{Phys. Rev. D} {\bfseries 108} (2023) 103542} [\href{https://arxiv.org/abs/2306.09456}{{\ttfamily 2306.09456}}].

\bibitem{Irsic:2023equ}
V.~Ir\v{s}i\v{c} et~al., \emph{{Unveiling dark matter free streaming at the smallest scales with the high redshift Lyman-alpha forest}}, \href{https://doi.org/10.1103/PhysRevD.109.043511}{\emph{Phys. Rev. D} {\bfseries 109} (2024) 043511} [\href{https://arxiv.org/abs/2309.04533}{{\ttfamily 2309.04533}}].

\bibitem{Delos:2023dwq}
M.S.~Delos, \emph{{An analytical description of substructure-induced gravitational perturbations in stellar systems}}, \href{https://doi.org/10.1093/mnras/stae715}{\emph{Mon. Not. Roy. Astron. Soc.} {\bfseries 529} (2024) 2349} [\href{https://arxiv.org/abs/2312.13338}{{\ttfamily 2312.13338}}].

\bibitem{Esteban:2023xpk}
I.~Esteban, A.H.G.~Peter and S.Y.~Kim, \emph{{Milky~Way satellite velocities reveal the dark matter power spectrum at small scales}}, \href{https://doi.org/10.1103/PhysRevD.110.123013}{\emph{Phys. Rev. D} {\bfseries 110} (2024) 123013} [\href{https://arxiv.org/abs/2306.04674}{{\ttfamily 2306.04674}}].

\bibitem{Nadler:2024ims}
E.O.~Nadler, V.~Gluscevic, T.~Driskell, R.H.~Wechsler, L.A.~Moustakas, A.~Benson et~al., \emph{{Forecasts for Galaxy Formation and Dark Matter Constraints from Dwarf Galaxy Surveys}}, \href{https://doi.org/10.3847/1538-4357/ad3bb1}{\emph{Astrophys. J.} {\bfseries 967} (2024) 61} [\href{https://arxiv.org/abs/2401.10318}{{\ttfamily 2401.10318}}].

\bibitem{Xiao:2024qay}
H.~Xiao, L.~Dai and M.~McQuinn, \emph{{Detecting dark matter substructures on small scales with fast radio bursts}}, \href{https://doi.org/10.1103/PhysRevD.110.023516}{\emph{Phys. Rev. D} {\bfseries 110} (2024) 023516} [\href{https://arxiv.org/abs/2401.08862}{{\ttfamily 2401.08862}}].

\bibitem{Ji:2024ott}
L.~Ji and L.~Dai, \emph{{Effects of Subhalos on Interpreting Highly Magnified Sources Near Lensing Caustics}},  \href{https://arxiv.org/abs/2407.09594}{{\ttfamily 2407.09594}}.

\bibitem{deKruijf:2024voc}
J.~de~Kruijf, E.~Vanzan, K.K.~Boddy, A.~Raccanelli and N.~Bartolo, \emph{{Searching for blue-tilted power spectra in the dark ages}}, \href{https://doi.org/10.1103/PhysRevD.111.063507}{\emph{Phys. Rev. D} {\bfseries 111} (2025) 063507} [\href{https://arxiv.org/abs/2408.04991}{{\ttfamily 2408.04991}}].

\bibitem{Lazare:2024uvj}
H.~Lazare, J.~Flitter and E.D.~Kovetz, \emph{{Constraints on the fuzzy dark matter mass window from high-redshift observables}}, \href{https://doi.org/10.1103/PhysRevD.110.123532}{\emph{Phys. Rev. D} {\bfseries 110} (2024) 123532} [\href{https://arxiv.org/abs/2407.19549}{{\ttfamily 2407.19549}}].

\bibitem{Buckley:2025zgh}
M.R.~Buckley, P.~Du, N.~Fernandez and M.J.~Weikert, \emph{{General Constraints on Isocurvature from the CMB and Ly-$\alpha$ Forest}},  \href{https://arxiv.org/abs/2502.20434}{{\ttfamily 2502.20434}}.

\bibitem{He:2025jwp}
A.~He, M.M.~Ivanov, S.~Bird, R.~An and V.~Gluscevic, \emph{{A Fresh Look at Neutrino Self-Interactions With the Lyman-$\alpha$ Forest: Constraints from EFT and PRIYA}},  \href{https://arxiv.org/abs/2503.15592}{{\ttfamily 2503.15592}}.

\bibitem{Narayanan:2000tp}
V.K.~Narayanan, D.N.~Spergel, R.~Dave and C.-P.~Ma, \emph{{Constraints on the mass of warm dark matter particles and the shape of the linear power spectrum from the Ly$\alpha$ forest}}, \href{https://doi.org/10.1086/317269}{\emph{Astrophys. J. Lett.} {\bfseries 543} (2000) L103} [\href{https://arxiv.org/abs/astro-ph/0005095}{{\ttfamily astro-ph/0005095}}].

\bibitem{Hansen:2001zv}
S.H.~Hansen, J.~Lesgourgues, S.~Pastor and J.~Silk, \emph{{Constraining the window on sterile neutrinos as warm dark matter}}, \href{https://doi.org/10.1046/j.1365-8711.2002.05410.x}{\emph{Mon. Not. Roy. Astron. Soc.} {\bfseries 333} (2002) 544} [\href{https://arxiv.org/abs/astro-ph/0106108}{{\ttfamily astro-ph/0106108}}].

\bibitem{Lewis:2002nc}
A.~Lewis and A.~Challinor, \emph{{Evolution of cosmological dark matter perturbations}}, \href{https://doi.org/10.1103/PhysRevD.66.023531}{\emph{Phys. Rev. D} {\bfseries 66} (2002) 023531} [\href{https://arxiv.org/abs/astro-ph/0203507}{{\ttfamily astro-ph/0203507}}].

\bibitem{Green:2003un}
A.M.~Green, S.~Hofmann and D.J.~Schwarz, \emph{{The power spectrum of SUSY - CDM on sub-galactic scales}}, \href{https://doi.org/10.1111/j.1365-2966.2004.08232.x}{\emph{Mon. Not. Roy. Astron. Soc.} {\bfseries 353} (2004) L23} [\href{https://arxiv.org/abs/astro-ph/0309621}{{\ttfamily astro-ph/0309621}}].

\bibitem{Viel:2005qj}
M.~Viel, J.~Lesgourgues, M.G.~Haehnelt, S.~Matarrese and A.~Riotto, \emph{{Constraining warm dark matter candidates including sterile neutrinos and light gravitinos with WMAP and the Lyman-alpha forest}}, \href{https://doi.org/10.1103/PhysRevD.71.063534}{\emph{Phys. Rev. D} {\bfseries 71} (2005) 063534} [\href{https://arxiv.org/abs/astro-ph/0501562}{{\ttfamily astro-ph/0501562}}].

\bibitem{Lesgourgues:2006nd}
J.~Lesgourgues and S.~Pastor, \emph{{Massive neutrinos and cosmology}}, \href{https://doi.org/10.1016/j.physrep.2006.04.001}{\emph{Phys. Rept.} {\bfseries 429} (2006) 307} [\href{https://arxiv.org/abs/astro-ph/0603494}{{\ttfamily astro-ph/0603494}}].

\bibitem{Viel:2007mv}
M.~Viel, G.D.~Becker, J.S.~Bolton, M.G.~Haehnelt, M.~Rauch and W.L.W.~Sargent, \emph{{How cold is cold dark matter? Small scales constraints from the flux power spectrum of the high-redshift Lyman-alpha forest}}, \href{https://doi.org/10.1103/PhysRevLett.100.041304}{\emph{Phys. Rev. Lett.} {\bfseries 100} (2008) 041304} [\href{https://arxiv.org/abs/0709.0131}{{\ttfamily 0709.0131}}].

\bibitem{Boyarsky:2008xj}
A.~Boyarsky, J.~Lesgourgues, O.~Ruchayskiy and M.~Viel, \emph{{Lyman-alpha constraints on warm and on warm-plus-cold dark matter models}}, \href{https://doi.org/10.1088/1475-7516/2009/05/012}{\emph{JCAP} {\bfseries 05} (2009) 012} [\href{https://arxiv.org/abs/0812.0010}{{\ttfamily 0812.0010}}].

\bibitem{Erickcek:2011us}
A.L.~Erickcek and K.~Sigurdson, \emph{{Reheating Effects in the Matter Power Spectrum and Implications for Substructure}}, \href{https://doi.org/10.1103/PhysRevD.84.083503}{\emph{Phys. Rev. D} {\bfseries 84} (2011) 083503} [\href{https://arxiv.org/abs/1106.0536}{{\ttfamily 1106.0536}}].

\bibitem{Lancaster:2017ksf}
L.~Lancaster, F.-Y.~Cyr-Racine, L.~Knox and Z.~Pan, \emph{{A tale of two modes: Neutrino free-streaming in the early universe}}, \href{https://doi.org/10.1088/1475-7516/2017/07/033}{\emph{JCAP} {\bfseries 07} (2017) 033} [\href{https://arxiv.org/abs/1704.06657}{{\ttfamily 1704.06657}}].

\bibitem{Irsic:2017ixq}
V.~Ir\v{s}i\v{c} et~al., \emph{{New Constraints on the free-streaming of warm dark matter from intermediate and small scale Lyman-$\alpha$ forest data}}, \href{https://doi.org/10.1103/PhysRevD.96.023522}{\emph{Phys. Rev. D} {\bfseries 96} (2017) 023522} [\href{https://arxiv.org/abs/1702.01764}{{\ttfamily 1702.01764}}].

\bibitem{wdm}
V.~Ir\v{s}i\v{c} et~al., \emph{{New Constraints on the free-streaming of warm dark matter from intermediate and small scale Lyman-$\alpha$ forest data}}, \href{https://doi.org/10.1103/PhysRevD.96.023522}{\emph{Phys. Rev. D} {\bfseries 96} (2017) 023522} [\href{https://arxiv.org/abs/1702.01764}{{\ttfamily 1702.01764}}].

\bibitem{Miller:2019pss}
C.~Miller, A.L.~Erickcek and R.~Murgia, \emph{{Constraining nonthermal dark matter\textquoteright{}s impact on the matter power spectrum}}, \href{https://doi.org/10.1103/PhysRevD.100.123520}{\emph{Phys. Rev. D} {\bfseries 100} (2019) 123520} [\href{https://arxiv.org/abs/1908.10369}{{\ttfamily 1908.10369}}].

\bibitem{Erickcek:2021fsu}
A.L.~Erickcek, P.~Ralegankar and J.~Shelton, \emph{{Cannibalism's lingering imprint on the matter power spectrum}}, \href{https://doi.org/10.1088/1475-7516/2022/01/017}{\emph{JCAP} {\bfseries 01} (2022) 017} [\href{https://arxiv.org/abs/2106.09041}{{\ttfamily 2106.09041}}].

\bibitem{Ballesteros:2020adh}
G.~Ballesteros, M.A.G.~Garcia and M.~Pierre, \emph{{How warm are non-thermal relics? Lyman-$\alpha$ bounds on out-of-equilibrium dark matter}}, \href{https://doi.org/10.1088/1475-7516/2021/03/101}{\emph{JCAP} {\bfseries 03} (2021) 101} [\href{https://arxiv.org/abs/2011.13458}{{\ttfamily 2011.13458}}].

\bibitem{Sarkar:2021pqh}
A.K.~Sarkar, K.L.~Pandey and S.K.~Sethi, \emph{{Using the redshift evolution of the Lyman-\ensuremath{\alpha} effective opacity as a probe of dark matter models}}, \href{https://doi.org/10.1088/1475-7516/2021/10/077}{\emph{JCAP} {\bfseries 10} (2021) 077} [\href{https://arxiv.org/abs/2101.09917}{{\ttfamily 2101.09917}}].

\bibitem{Garcia:2023qab}
M.A.G.~Garcia, M.~Pierre and S.~Verner, \emph{{New window into gravitationally produced scalar dark matter}}, \href{https://doi.org/10.1103/PhysRevD.108.115024}{\emph{Phys. Rev. D} {\bfseries 108} (2023) 115024} [\href{https://arxiv.org/abs/2305.14446}{{\ttfamily 2305.14446}}].

\bibitem{Liu:2024pjg}
R.~Liu, W.~Hu and H.~Xiao, \emph{{Warm and fuzzy dark matter: Free streaming of wave dark matter}}, \href{https://doi.org/10.1103/PhysRevD.111.023535}{\emph{Phys. Rev. D} {\bfseries 111} (2025) 023535} [\href{https://arxiv.org/abs/2406.12970}{{\ttfamily 2406.12970}}].

\bibitem{Liu:2025lts}
R.~Liu, W.~Hu and H.~Xiao, \emph{{Interference with Gravitational Instability: Hot and Fuzzy Dark Matter}},  \href{https://arxiv.org/abs/2504.01937}{{\ttfamily 2504.01937}}.

\bibitem{Dai:2019lud}
L.~Dai and J.~Miralda-Escud\'e, \emph{{Gravitational Lensing Signatures of Axion Dark Matter Minihalos in Highly Magnified Stars}}, \href{https://doi.org/10.3847/1538-3881/ab5e83}{\emph{Astron. J.} {\bfseries 159} (2020) 49} [\href{https://arxiv.org/abs/1908.01773}{{\ttfamily 1908.01773}}].

\bibitem{Ramani:2020hdo}
H.~Ramani, T.~Trickle and K.M.~Zurek, \emph{{Observability of Dark Matter Substructure with Pulsar Timing Correlations}}, \href{https://doi.org/10.1088/1475-7516/2020/12/033}{\emph{JCAP} {\bfseries 12} (2020) 033} [\href{https://arxiv.org/abs/2005.03030}{{\ttfamily 2005.03030}}].

\bibitem{Lee:2020wfn}
V.S.H.~Lee, A.~Mitridate, T.~Trickle and K.M.~Zurek, \emph{{Probing Small-Scale Power Spectra with Pulsar Timing Arrays}}, \href{https://doi.org/10.1007/JHEP06(2021)028}{\emph{JHEP} {\bfseries 06} (2021) 028} [\href{https://arxiv.org/abs/2012.09857}{{\ttfamily 2012.09857}}].

\bibitem{Blinov:2021axd}
N.~Blinov, M.J.~Dolan, P.~Draper and J.~Shelton, \emph{{Dark Matter Microhalos From Simplified Models}}, \href{https://doi.org/10.1103/PhysRevD.103.103514}{\emph{Phys. Rev. D} {\bfseries 103} (2021) 103514} [\href{https://arxiv.org/abs/2102.05070}{{\ttfamily 2102.05070}}].

\bibitem{Lee:2021zqw}
V.S.H.~Lee, S.R.~Taylor, T.~Trickle and K.M.~Zurek, \emph{{Bayesian Forecasts for Dark Matter Substructure Searches with Mock Pulsar Timing Data}}, \href{https://doi.org/10.1088/1475-7516/2021/08/025}{\emph{JCAP} {\bfseries 08} (2021) 025} [\href{https://arxiv.org/abs/2104.05717}{{\ttfamily 2104.05717}}].

\bibitem{Delos:2021rqs}
M.S.~Delos and T.~Linden, \emph{{Dark matter microhalos in the solar neighborhood: Pulsar timing signatures of early matter domination}}, \href{https://doi.org/10.1103/PhysRevD.105.123514}{\emph{Phys. Rev. D} {\bfseries 105} (2022) 123514} [\href{https://arxiv.org/abs/2109.03240}{{\ttfamily 2109.03240}}].

\bibitem{Graham:2024hah}
P.W.~Graham and H.~Ramani, \emph{{Constraints on dark matter from dynamical heating of stars in ultrafaint dwarfs. II. Substructure and the primordial power spectrum}}, \href{https://doi.org/10.1103/PhysRevD.110.075012}{\emph{Phys. Rev. D} {\bfseries 110} (2024) 075012} [\href{https://arxiv.org/abs/2404.01378}{{\ttfamily 2404.01378}}].

\bibitem{Salehian:2020bon}
B.~Salehian, M.H.~Namjoo and D.I.~Kaiser, \emph{{Effective theories for a nonrelativistic field in an expanding universe: Induced self-interaction, pressure, sound speed, and viscosity}}, \href{https://doi.org/10.1007/JHEP07(2020)059}{\emph{JHEP} {\bfseries 07} (2020) 059} [\href{https://arxiv.org/abs/2005.05388}{{\ttfamily 2005.05388}}].

\bibitem{Salehian:2021khb}
B.~Salehian, H.-Y.~Zhang, M.A.~Amin, D.I.~Kaiser and M.H.~Namjoo, \emph{{Beyond Schr\"odinger-Poisson: nonrelativistic effective field theory for scalar dark matter}}, \href{https://doi.org/10.1007/JHEP09(2021)050}{\emph{JHEP} {\bfseries 09} (2021) 050} [\href{https://arxiv.org/abs/2104.10128}{{\ttfamily 2104.10128}}].

\bibitem{Levkov:2018kau}
D.G.~Levkov, A.G.~Panin and I.I.~Tkachev, \emph{{Gravitational Bose-Einstein condensation in the kinetic regime}}, \href{https://doi.org/10.1103/PhysRevLett.121.151301}{\emph{Phys. Rev. Lett.} {\bfseries 121} (2018) 151301} [\href{https://arxiv.org/abs/1804.05857}{{\ttfamily 1804.05857}}].

\bibitem{Dodelson:2020bqr}
S.~Dodelson and F.~Schmidt, \emph{{Modern Cosmology}}, Academic Press (2020), \href{https://doi.org/10.1016/C2017-0-01943-2}{10.1016/C2017-0-01943-2}.

\bibitem{Baumann:2022mni}
D.~Baumann, \emph{{Cosmology}}, Cambridge University Press (7, 2022), \href{https://doi.org/10.1017/9781108937092}{10.1017/9781108937092}.

\bibitem{Brandenberger:1987}
R.~Brandenberger, N.~Kaiser and N.~Turok, \emph{Dissipationless clustering of neutrinos around a cosmic-string loop}, \href{https://doi.org/10.1103/PhysRevD.36.2242}{\emph{Phys. Rev. D} {\bfseries 36} (1987) 2242}.

\bibitem{Capanelli:2025nrj}
C.~Capanelli, W.~Hu and E.~McDonough, \emph{{Wave Interference in Self-Interacting Fuzzy Dark Matter}},  \href{https://arxiv.org/abs/2503.21865}{{\ttfamily 2503.21865}}.

\bibitem{Hu:2000ke}
W.~Hu, R.~Barkana and A.~Gruzinov, \emph{{Cold and fuzzy dark matter}}, \href{https://doi.org/10.1103/PhysRevLett.85.1158}{\emph{Phys. Rev. Lett.} {\bfseries 85} (2000) 1158} [\href{https://arxiv.org/abs/astro-ph/0003365}{{\ttfamily astro-ph/0003365}}].

\bibitem{May:2021wwp}
S.~May and V.~Springel, \emph{{Structure formation in large-volume cosmological simulations of fuzzy dark matter: impact of the non-linear dynamics}}, \href{https://doi.org/10.1093/mnras/stab1764}{\emph{Mon. Not. Roy. Astron. Soc.} {\bfseries 506} (2021) 2603} [\href{https://arxiv.org/abs/2101.01828}{{\ttfamily 2101.01828}}].

\bibitem{May:2022gus}
S.~May and V.~Springel, \emph{{The halo mass function and filaments in full cosmological simulations with fuzzy dark matter}}, \href{https://doi.org/10.1093/mnras/stad2031}{\emph{Mon. Not. Roy. Astron. Soc.} {\bfseries 524} (2023) 4256} [\href{https://arxiv.org/abs/2209.14886}{{\ttfamily 2209.14886}}].

\bibitem{Planck:2015fie}
{\scshape Planck} collaboration, \emph{{Planck 2015 results. XIII. Cosmological parameters}}, \href{https://doi.org/10.1051/0004-6361/201525830}{\emph{Astron. Astrophys.} {\bfseries 594} (2016) A13} [\href{https://arxiv.org/abs/1502.01589}{{\ttfamily 1502.01589}}].

\bibitem{Planck:2018vyg}
{\scshape Planck} collaboration, \emph{{Planck 2018 results. VI. Cosmological parameters}}, \href{https://doi.org/10.1051/0004-6361/201833910}{\emph{Astron. Astrophys.} {\bfseries 641} (2020) A6} [\href{https://arxiv.org/abs/1807.06209}{{\ttfamily 1807.06209}}].

\bibitem{Delos:2023eve}
M.S.~Delos, \emph{{Accurate halo mass functions from the simplest excursion set theory}}, \href{https://doi.org/10.1093/mnras/stae141}{\emph{Mon. Not. Roy. Astron. Soc.} {\bfseries 528} (2024) 1372} [\href{https://arxiv.org/abs/2311.17986}{{\ttfamily 2311.17986}}].

\bibitem{Gorghetto:2022sue}
M.~Gorghetto, E.~Hardy, J.~March-Russell, N.~Song and S.M.~West, \emph{{Dark Photon Stars: Formation and Role as Dark Matter Substructure}},  \href{https://arxiv.org/abs/2203.10100}{{\ttfamily 2203.10100}}.

\bibitem{Chang:2024fol}
J.H.~Chang, P.J.~Fox and H.~Xiao, \emph{{Axion stars: mass functions and constraints}}, \href{https://doi.org/10.1088/1475-7516/2024/08/023}{\emph{JCAP} {\bfseries 08} (2024) 023} [\href{https://arxiv.org/abs/2406.09499}{{\ttfamily 2406.09499}}].

\bibitem{Gorghetto:2024vnp}
M.~Gorghetto, E.~Hardy and G.~Villadoro, \emph{{More axion stars from strings}}, \href{https://doi.org/10.1007/JHEP08(2024)126}{\emph{JHEP} {\bfseries 08} (2024) 126} [\href{https://arxiv.org/abs/2405.19389}{{\ttfamily 2405.19389}}].

\bibitem{Gorghetto:2020qws}
M.~Gorghetto, E.~Hardy and G.~Villadoro, \emph{{More axions from strings}}, \href{https://doi.org/10.21468/SciPostPhys.10.2.050}{\emph{SciPost Phys.} {\bfseries 10} (2021) 050} [\href{https://arxiv.org/abs/2007.04990}{{\ttfamily 2007.04990}}].

\bibitem{Meszaros}
P.~{Mészáros}, \emph{{The behaviour of point masses in an expanding cosmological substratum.}}, {\emph{\aap} {\bfseries 37} (1974) 225}.

\bibitem{Woo:2008nn}
T.-P.~{Woo} and T.~{Chiueh}, \emph{{High Resolution Simulation on Structure Formation with Extremely Light Bosonic Dark Matter}}, \href{https://doi.org/10.1088/0004-637X/697/1/850}{\emph{Astrophys. J.} {\bfseries 697} (2009) 850} [\href{https://arxiv.org/abs/0806.0232}{{\ttfamily 0806.0232}}].

\end{thebibliography}\endgroup

\clearpage
\appendix

\section{Appendix}
\subsection{Notation \& Conventions}
\label{sec:notconv}
The background Friedmann–Lemaître–Robertson–Walker spacetime is given by
\beq
    \dl s^2 = -\dl t^2 + a(t)^2 \dl\bx \cdot \dl\bx \,,
\eeq
where $a(t)$ is the scale factor, $\bx$ are comoving coordinates, and we set $c=1$.
Generally, we use comoving lengths and coordinates unless stated otherwise, e.\,g.\ $\bx = a(t)^{-1} \mathbf{r}$, where $\mathbf{r}$ is in proper (``physical") length units.

Our Fourier conventions for finite volumes $V$ and infinite volumes, respectively, are
\beq
    f(\bx) &= V^{-1}\sum_\bk e^{i\bk \cdot \bx} f_\bk,
    &\!\!
    f_\bk &= \int_V \! \dl\bx\, e^{-i\bk \cdot \bx} f(\bx),
    &\!\!
    \int_V \! \dl\bx\, e^{\pm i(\bk+\bk') \cdot \bx} &= V\delta_{\bk-\bk'} \,,
    \\
    f(\bx) &= \int \! \frac{\dl\bk}{(2\pi)^3} e^{i\bk \cdot \bx} f(\bk),
    &\!\!
    f(\bk) &= \int \! \dl\bx\, e^{-i\bk \cdot \bx} f(\bx),
    &\!\!
    \int \! \dl\bx\, e^{\pm i(\bk+\bk')\cdot \bx} &= (2\pi)^3\delta_{\mathrm{D}}(\bk+\bk') \,.
\eeq
Here $\delta_{\bk-\bk'}$ is the Kronecker $\delta$, and $\delta_{\mathrm{D}}$ is the Dirac $\delta$-distribution.
Note that the finite- and infinite-volume cases are linked by:
\beq
    \label{eq:finite_infinite}
    &\qquad\int_\bk \quad\longleftrightarrow\quad \int \frac{\dl\bk}{(2\pi)^3}\quad\longleftrightarrow\quad \frac{1}{V}\sum_\bk\,,\\
    &\ddelta(\bk+\bk')\longleftrightarrow (2\pi)^3 \delta_{\mathrm{D}}(\bk+\bk') \longleftrightarrow V\delta_{\bk,-\bk'}\,,
\eeq
and formally $V\leftrightarrow (2\pi)^3\delta_{\mathrm{D}}(\mathbf{0})$.
The first entry on each line of \cref{eq:finite_infinite} represents the shorthand notation
\beq
    f(\bx) &= \int_\bk e^{i\bk \cdot \bx}f_\bk,\qquad f_\bk=\int_\bx e^{-i\bk \cdot \bx}f(\bx),\quad \int_\bx e^{\pm i(\bk+\bk')\cdot \bx}=\ddelta(\bk+\bk'),
\eeq
where the integral without the explicit measure can stand in for finite- or infinite-volume cases.

For a statistically homogeneous field $f$, the power spectrum $P_f$ is defined:
\beq
    \langle f(\bk)f(\bk')\rangle =(2\pi)^3 P_f(\bk)\delta_{\mathrm{D}}(\bk+\bk') \,,
    \qquad
    \langle f_\bk f_{\bk'}\rangle = V P_f(\bk)\delta_{\bk-\bk'}.
\eeq
The angled brackets stand for the average over the ensemble.

\subsection{Summary of Power Spectrum Evolution}
\label{sec:summaryPS}
The final result for the power spectrum evolution can be written as follows:\footnote{This definition of the adiabatic transfer function is consistent with one in \cite{Amin:2025dtd}, but differs from our main text. There, it is defined as $[\mathcal{T}^{(\mathrm{ad})}_k(y,y_0)]^2 = P_\delta(y,k) / P_\zeta(k)$, consistent with the broader literature.}
\beq
    \label{eq:MainResultPS}
    P_{\delta}(y,k) &= \underbrace{{P}^{(\mathrm{ad})}_{\delta}(y_0,k) \ml[\mathcal{T}_k^{(\mathrm{ad})}(y,y_0)\mr]^2 \vphantom{\int_{y_0}^y}}_{\text{adiabatic IC + evolution}} \!\! +
    \underbrace{P^{(\mathrm{iso})}_\delta(y_0,k) \ml[1 + 3 \int_{y_0}^y \! \frac{\dl y'}{\sqrt{1+y'}} \mathcal{T}^{(\mathrm{b})}_k(y, y')\mathcal{T}^{(\mathrm{c})}_k(y,y')\mr]}_{\text{isocurvature IC + evolution}},
\eeq
where $y = a/a_{\mathrm{eq}}$ and $y_0$ is at an initial time when all wavenumber-$k$ modes of interest are sub-horizon, and the field modes of interest are non-relativistic; the initial conditions (IC) are specified at that time. The three different $\mathcal{T}_k^{(\mathrm{ad}, \mathrm{a}, \mathrm{c})}$ in the above expressions are growth functions, which describe how density perturbations evolve due to gravitational clustering and free streaming.
They are given by:
\beq
    \mathcal{T}^{\mathrm{ad}}_k(y,y_0) &= \mathcal{T}^{(\mathrm{a})}_k(y,y_0) + \frac{1}{2} \frac{\dl\ln(P^{({\mathrm{ad}})}_{\delta}(y_0,k))}{\dl\ln(y_0)} \sqrt{1+y_0} \, \mathcal{T}^{(\mathrm{b})}_k(y,y_0),
    \\
    \mathcal{T}^{(\mathrm{a}, \mathrm{b}, \mathrm{c})}_k(y,y') &= T^{\mathrm{fs}\,(\mathrm{a}, \mathrm{b}, \mathrm{c})}_k(y,y')+\frac{3}{2}\int_{y'}^y \frac{\dl y''}{\sqrt{1+y''}}\mathcal{T}^{(\mathrm{b})}_k(y,y'') T^{\mathrm{fs}\,(\mathrm{a}, \mathrm{b}, \mathrm{c})}_k(y'',y').\\
\eeq
The free-streaming kernels can be calculated based on an (isotropic) initial field power spectrum
\beq
    T^{\mathrm{fs}\,(\mathrm{a})}_k(y,y') &= \cos[\gamma\alpha_k^2\mathcal{F}(y,y')]\int_\bq f_0(q) \exp\ml[-i\hat{\bq} \cdot \hat{\bk}\frac{q}{k_*}\alpha_k\mathcal{F}(y,y')\mr],
    \\
    T^{\mathrm{fs}\,(\mathrm{b})}_k(y,y') &= \frac{1}{\gamma \alpha_k^2} \sin[\gamma \alpha_k^2\mathcal{F}(y,y')] \int_\bq f_0(q) \exp\ml[-i\hat{\bq}\cdot \hat{\bk}\frac{q}{k_*}\alpha_k\mathcal{F}(y,y')\mr],
    \\
    \hat{T}^{\mathrm{fs}\,(\mathrm{c})}_k(y,y') &= \frac{\int_\bq  f_0(|\bq+\bk/2|)f_0(|\bq-\bk/2|) \exp\ml[-i\hat{\bq}\cdot \hat{\bk}\frac{q}{k_*}\alpha_k \mathcal{F}(y,y')\mr]}{\int_\bq  f_0(|\bq+\bk/2|)f_0(|\bq-\bk/2|)}.
\eeq
Here, $\mathcal{F}(y,y') = \ln\ml[(y/y')(1+\sqrt{1+y'})^2 / (1+\sqrt{1+y})^2\mr]$ captures the functional dependence of the comoving distance traveled by a field fluctuation during the time interval between $y'$ and $y$, and the parameters
\beq
    \alpha_k \equiv \sqrt{2}\frac{k}{k_{\mathrm{eq}}}\frac{k_*}{a_{\mathrm{eq}} m},
    \qquad
    \gamma \equiv \frac{1}{2\sqrt{2}}\frac{a_{\mathrm{eq}} m}{k_*}\frac{k_{\mathrm{eq}}}{k_*},
\eeq
in the above equations.

The quantity $\alpha_k$ is a scaled version of the wavenumber. It is useful because for $\gamma \rightarrow 0$, the evolution of the isocurvature density power spectrum is universal in terms of $\alpha_k$ for all $k_*$ and $m$. The dimensionless variable $\gamma \ne 0$ captures wave-dynamical effects around the de Broglie scale which are not captured in the quasi-particle picture \cite{Amin:2025dtd}. In particular, such deviations in the power spectrum evolution occur at $\gamma\alpha_k \sim 1$ (equivalently $k\sim k_*$). Furthermore, $T^{(\mathrm{c})}$ appearing here is replaced by $T^{(a)}$ in the quasi-particle picture.

\begin{figure}
    \centering
    \includegraphics[width=0.75\linewidth]{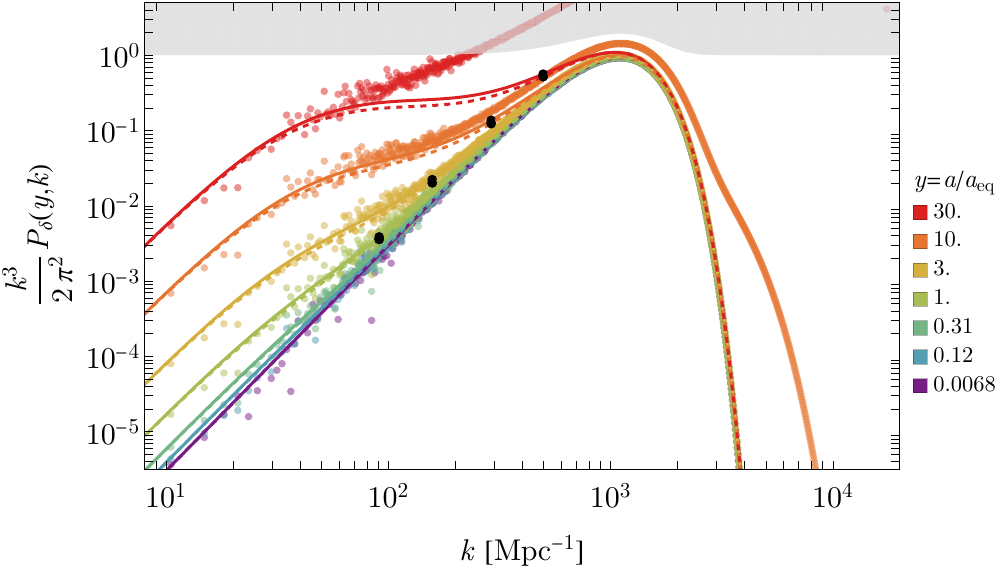}

    \caption{A comparison between the analytically predicted evolution of the isocurvature matter power spectrum and results from Schrödinger–Poisson simulations. The solid lines include wave effects. Dashed lines are based on the quasi-particle calculation (apart from the initial power spectrum). The above plot is for Maxwell–Boltzmann initial field spectra. The difference with the quasi-particle case appears at $k\ll k_*$, indicating the effect of $T^{\mathrm{fs}\,(\mathrm{c})} \ne T^{\mathrm{fs}\,(\mathrm{a})}$ even when $\gamma$ is small. Note that the for the above case, $\gamma \approx 0.08 \ll 1$, which restricts the impact of the sine and cosine appearing in the free-streaming kernels.}
    \label{fig:ParticleVsWaveGaussian}
\end{figure}

\subsection{Worked Examples}
\label{sec:examples}
Let us consider a simple form for the initial power spectra of the field $f_0(p) = (m/a^3\bar{\rho})P_\psi(p)$. If the initial $f_0$ is a top-hat function (uniform sphere) in momentum, then
\beq
    f_0(p) &= \frac{6\pi^2}{k_*^3} \Theta(k_* - p),\quad
    T^{\mathrm{fs}}_k =\frac{3}{x^3} \left(\sin(x) - x \cos(x)\right),
    \quad \text{with} \quad
    x = \frac{k_*}{m}k\Delta\eta,
    \\
    T^{\mathrm{fs}\,(\mathrm{a})}_k &= \cos\ml(\frac{k}{2k_*}x\mr) T_{\mathrm{fs}} \,,
    \quad
    T^{{\mathrm{fs}}\,(\mathrm{b})}_k = \Delta\eta\frac{\sin\ml(\frac{k}{2k_*}x\mr)}{\frac{k}{2k_*}x}T^{\mathrm{fs}}_k \,,
    \\
    \hat{T}^{\mathrm{fs}\,(\mathrm{c})}_k &= \frac{3}{x^3}\frac{\sin
        \ml[\left(1-\frac{k}{2
        k_*}\right) x\mr] + x
        \left\{\frac{k}{2 k_*}-\cos
        \ml[\left(1-\frac{k}{2
        k_*}\right)
        x\mr]\right\}}{\left
        (\frac{k}{4 k_*}+1\right)
        \left(\frac{k}{2
        k_*}-1\right)^2}\Theta(2k_*-k).
\eeq
In the limit that $k_*\rightarrow \infty$, but at fixed $k_*/m$ we get $T^{\mathrm{fs}\,(\mathrm{a}, \mathrm{c})} \rightarrow T^{\mathrm{fs}}$ and $T^{\mathrm{fs}\,(\mathrm{b})} \rightarrow \Delta \eta T^{\mathrm{fs}}$. This leads to expressions identical to the quasi-particle case in \cite{Amin:2025dtd}.

The initial isocurvature density power spectrum for this case is
\beq
    P_{\delta}^{(\mathrm{iso})}(\eta_0,k)=\frac{6 \pi ^2}{k_*^3} \left(1+\frac{k}{4
    k_*}\right) \left(1-\frac{k}{2
    k_*}\right)^2\Theta\ml(1-\frac{k}{2k_*}\mr).
\eeq
For $k\ll k_*$, the above $P_\delta^{(\mathrm{iso})}(\eta_0,k) = 2\pi^2/k_{\mathrm{wn}}^3$, with $k_{\mathrm{wn}} = 3^{-1/3}k_*$. The Jeans and free-streaming scales are determined by
\beq
    \sigma(\eta) = \frac{1}{\sqrt{5}}\frac{k_*}{a m}.
\eeq
For the case where $f_0(p)$ is a Gaussian (Maxwell–Boltzmann) distribution:
\beq
    &f_0(p)=\frac{(2\pi)^{3/2}}{k_*^3}e^{-\frac{p^2}{2k_*^2}},
    \quad
    T^{\mathrm{fs}}_k = e^{-\frac{x^2}{2}},\quad \text{with} \quad
    x = \frac{k_*}{m}k\Delta\eta,
    \\
    &T^{\mathrm{fs}\,(\mathrm{a})}_k = \cos\ml(\frac{k}{2k_*}x\mr) T^{\mathrm{fs}}_k \,,
    \quad
    T^{\mathrm{fs}\,(\mathrm{b})}_k = \Delta\eta\frac{\sin\ml(\frac{k}{2k_*}x\mr)}{\frac{k}{2k_*}x} T^{\mathrm{fs}}_k \,,
    \quad
    T^{\mathrm{fs}\,(\mathrm{c})}_k = e^{-\frac{x^2}{4}},
    \\
    &P_\delta^{(\mathrm{iso})}(\eta_0,k) = \frac{\pi^{3/2}}{k_*^3}e^{-\frac{k^2}{4k_*^2}},
    \quad
    \sigma(\eta) = \frac{k_*}{a m}.
\eeq
Note that for $k \ll k_*$, $P^{(\mathrm{iso})}_\delta = \pi^{3/2}k_*^{-3}$, with $k_{\mathrm{wn}}= (4\pi)^{1/6}k_*\approx 1.5 k_*$. In the limit $k_*\rightarrow \infty$, at fixed $k_*/m$, we get $T^{\mathrm{fs}\,(\mathrm{a})} \rightarrow T^{\mathrm{fs}}$, $T^{\mathrm{fs}\,(\mathrm{b})} \rightarrow T^{\mathrm{fs}}$. However, $T^{\mathrm{fs}\,(\mathrm{c})} \rightarrow \sqrt{T^{\mathrm{fs}}}\ne T^{{\mathrm{fs}}\,(\mathrm{a})}$. This is a consequence of wave interference effects (absent in the quasi-particle limit) discussed in \cref{sec:waveeffects}.

We provide a comparison of the numerically calculated power spectrum and our predictions in \cref{fig:ParticleVsWaveGaussian}. For these plots we chose to match $k_{\mathrm{wn}}$, which determined the amplitude of the white noise part of the initial spectrum, as $k_{\mathrm{J}}^{\mathrm{eq}}$ to the uniform sphere case discussed in the main text. This required $k_* = \SI{0.455e3}{\per\Mpc}$ and $m = \SI{1.02e-19}{\eV}$, corresponding to $\gamma = 0.08$. The difference between power spectra with and without wave-dynamical effects ($\gamma \rightarrow 0$, $T^{\mathrm{fs}\,(\mathrm{c})} \rightarrow T^{\mathrm{fs}\,(\mathrm{a})}$) is shown in \cref{fig:ParticleVsWaveGaussian}.

\subsection{Details of the Numerical Simulations}
\label{sec:NumericalDetails}
In this appendix, we provide details of both how to set up initial conditions for the field which includes both adiabatic and isocurvature density perturbations, as well as the time evolution of the fields thereafter.

\subsubsection{Initial Conditions}
\label{sec:AlgoIC}
We here provide an algorithm for generating initial field configurations $\psi(\bx) = \psi^{\mathrm{R}}(\bx) + i\psi^{\mathrm{I}}(\bx)$ (where $\psi^{\mathrm{R}} = \Re(\psi)$ and $\psi^{\mathrm{I}} = \Im(\psi)$) given some
\beq
    f_0(q) = \left(\frac{m}{a^3\bar{\rho}}\right)P_{\psi}(q)=\left(\frac{m}{a^3\bar{\rho}}\right)\left[P_{\psi^{\mathrm{R}}}(q) + P_{\psi^{\mathrm{I}}}(q)\right],
\eeq
with $P_{\psi^{\mathrm{R}}} = P_{\psi^{\mathrm{I}}}$. All quantities in this subsection are evaluated at the initial time. The procedure described here is a concrete implementation to numerically obtain the initial field shown in \cref{eq:psi-ic}.

The field configurations must be such that the density perturbation power spectra are consistent with
\beq
    \label{eq:PS_delta_IC}
    P_\delta(k) \approx P_\delta^{(\mathrm{ad})}(k) + P_\delta^{({\mathrm{iso}})}(k).
\eeq
Here, $P_\delta^{(\mathrm{iso})}$ is related to $f_0$ through \cref{eq:Piso_initial} and the initial adiabatic spectrum is\footnote{This spectrum has ignored free-streaming effects from production to $y_0$, and we have also
assumed that all scales of interest are sufficiently sub-horizon at this initial time.  Using this adiabatic spectrum we first generate a density field realization $\delta_{\mathrm{ad}}(\bk)$. A derivation of this expression can be found in, for example, \cite{Baumann:2022mni}. In our main text $\mathcal{R} \rightarrow \zeta$.}
\beq
    \frac{k^3}{2\pi^2}P^{(\mathrm{ad})}_{\delta}(k)\approx 36{\Delta}^2_{\mathcal{R}}(k)\left[3+\ln\ml(0.15\frac{k}{k_{\mathrm{eq}}}\mr) - \ln\ml(\frac{4}{y_0}\mr)\right]^2,
\eeq
with $\Delta^2_{\mathcal{R}}(k)$ being the inflationary curvature spectrum obtained e.\,g.\ from Planck (2015) \cite{Planck:2015fie}, and $y_0 = a(\eta_0)/a_{\mathrm{eq}}$ is specified by the initial time when we begin our simulations. Note that $P_\delta^{(\mathrm{ad})}$ is the leading contribution to $P_\delta$ for $k\lesssim 10^{-2}k_*$. In this regime, we want $P_\delta\approx P_{\delta}^{(\mathrm{ad})}$ (and similarly for the time derivative of the power spectrum, which is related to the bulk velocity perturbations).

To this end, using $P^{(\mathrm{ad})}_\delta$, we generate a realization $\delta^{\mathrm{ad}}_\bk$ in Fourier space, and correspondingly, $\delta_{\mathrm{ad}}(\bx)$ in position space. The bulk velocity perturbation $\bv^{\mathrm{ad}}_\bk$ (and $\bv_{\mathrm{ad}}(\bx)$) follows from the fluid continuity equation:
\beq
    \bv^{\mathrm{ad}}_\bk \approx i \frac{\bk}{k}\frac{k_{\mathrm{eq}}}{k}\frac{\delta^{\mathrm{ad}}_\bk}{\sqrt{2}y_0 \ln\ml[(3y_0/4)(k/k_{\mathrm{eq}})\mr]}.
\eeq
These $\delta_{\mathrm{ad}}(\bx)$ and $\bv_{\mathrm{ad}}(\bx)$ are then used to generated the desired initial field realization. In the algorithm below, the $\langle\dots\rangle$ means average over an ensemble of the field generated by drawing the initial field power spectrum, but not over the $\delta_{\mathrm{ad}}$ and $\bv_{\mathrm{ad}}$. The velocity field corresponding to $\psi$ is given by $\bv = (a^3\bar{\rho}/m) a^{-1} \left[\psi^{\mathrm{R}} \nabla \psi^{\mathrm{I}} - \{\mathrm{R} \leftrightarrow \mathrm{I}\}\right]$.

The steps for generating the desired initial field configuration are as follows:
\begin{enumerate}
    \item \label{item:algo-ic-1}
    Create fields $\hat{\psi}^{\mathrm{R},\mathrm{I}}(\bx)$ such that $\langle \hat{\psi}^{\mathrm{R}, \mathrm{I}}(\bx) \hat{\psi}^{\mathrm{R}, \mathrm{I}}(\bx')\rangle = (1/2) [1 + \delta_{\mathrm{ad}}(\bx)] \delta_{\mathrm{D}}(\bx - \bx') \delta_{\mathrm{R}, \mathrm{I}}$. That is, at each point in position space the field value is independently drawn from a Gaussian distribution of field amplitudes with variance given by $(1/2)[1+\delta_{\mathrm{ad}}(\bx)]$ and with zero mean.
    \item \label{item:algo-ic-2}
    Fourier transform these fields to obtain $\hat{\psi}^{\mathrm{R},\mathrm{I}}_{\bq}$.
    \item \label{item:algo-ic-3}
    Multiply by $\sqrt{{P}_{\psi^{\mathrm{R},\mathrm{I}}}(q)}$ to obtain $\tilde{\psi}^{\mathrm{R},\mathrm{I}}_\bq = \sqrt{{P}_{\psi^{\mathrm{R},\mathrm{I}}}(q)}\hat{\psi}^{\mathrm{R},\mathrm{I}}_{\bq}$.
    \item Fourier transform back to position space to obtain $\tilde{\psi}^{\mathrm{R},\mathrm{I}}(\bx)$, with the total complex field $\tilde{\psi} = \tilde{\psi}^{\mathrm{R}} + i\tilde{\psi}^{\mathrm{I}}$. This field will have the correct ensemble-averaged density perturbations $\langle\tilde{\delta}(\bx)\rangle = \delta_{\mathrm{ad}}(\bx)$. However, $\langle \tilde{\bv}(\bx)\rangle = 0$. We rectify this in the next two steps.
    \item Solve $\nabla^2 S_{\mathrm{ad}} = a^2 \, \nabla \cdot \bv_{\mathrm{ad}}$ to obtain $S_{\mathrm{ad}}(\bx)$.
    \item Rotate the field $\tilde{\psi}(y_0,\bx)$:
    $\psi = e^{iS_{\mathrm{ad}}} \tilde{\psi}$. This new rotated field $\psi(y_0, \bx)$ has the correct properties $\langle\delta(y_0,\bx)\rangle = \delta_{\mathrm{ad}}(y_0, \bx)$ and $\langle\bv(y_0, \bx)\rangle = \bv_{\mathrm{ad}}(y_0, \bx)$.
\end{enumerate}

To see why the algorithm achieves the desired result, note that using \ref{item:algo-ic-1}., \ref{item:algo-ic-2}.\ and \ref{item:algo-ic-3}.\ in a finite box of volume $V = L^3$, $\langle\tilde{\psi}^{\mathrm{R}}_\bq \tilde{\psi}^{*\mathrm{R}}_\bp\rangle = P_{\psi^{\mathrm{R}}}(q) V \delta_{\bp, \bq} + \sqrt{P_{\psi^{\mathrm{R}}}(q){P}_{\psi^{\mathrm{R}}}(p)} \, \delta^{\mathrm{ad}}_{\bq-\bp}$ (and similarly for $\mathrm{R} \rightarrow \mathrm{I}$). That is, for $\bp = \bq$, we recover $\langle |\psi^{\mathrm{R}}_\bq|^2\rangle = P_{\psi^{\mathrm{R}}}(q)V$ since $\delta^{\mathrm{ad}}_{\bq - \bp = 0} = 0$ by definition. So the field amplitudes are specified entirely by the power spectrum of the field in spite of the spatially dependent variance. As the next check, note that
\beq
    (a^3\bar{\rho})\langle \bv\rangle 
    &=
    a^{-1} \langle \psi^\mathrm{R} \nabla\psi^{\mathrm{I}} - \{\mathrm{R} \leftrightarrow \mathrm{I}\}\rangle,
    \\
    &= \left\langle \left[\cos(S_{\mathrm{ad}}) \tilde{\psi}^{\mathrm{R}} - \sin(S_{\mathrm{ad}}) \tilde{\psi}^{\mathrm{I}}\right] \nabla \left[\sin(S_{\mathrm{ad}}) \tilde{\psi}^{\mathrm{R}} + \cos(S_{\mathrm{ad}}) \tilde{\psi}^{\mathrm{I}}\right] - \{\mathrm{R} \leftrightarrow \mathrm{I}\}\right\rangle,
    \\
    &=(a^3\bar{\rho})\bv_{\mathrm{ad}} \,.
\eeq
where we used $\langle\tilde{\bv}\rangle = 0$. With $\langle \bv\rangle = \bv_{\mathrm{ad}}$ and $\langle\delta\rangle = \delta_{\mathrm{ad}}$, our initial field configuration will be consistent with the desired density power spectrum in \cref{eq:PS_delta_IC}.

The algorithm above is based on and is consistent with the relativistic version provided in \cite{Ling:2024qfv}. However, in \cite{Ling:2024qfv} since the adiabatic perturbation was outside the horizon, we made our field configurations consistent with zero adiabatic velocity perturbation (in an ensemble-averaged sense). Here, we are focused on sub-horizon scales, and we allow for  velocity perturbations to be present.

We note that the gravitational potentials due to radiation are ignored on sub-horizon scales since they are expected to decay after horizon entry. That is, we are in a regime similar to the applicability of the Mészáros equation \cite{Meszaros}.

\subsubsection{Time Evolution}
\label{sec:AxiREPO}
We will briefly review the numerical methods implemented in \code{AxiREPO} (also see \cite{May:2021wwp, May:2022gus}), as relevant for our simulations.
The simulation volume consists of a cubic box of side length $L$ with periodic boundary conditions, which is intended to sample the matter distribution in the universe.
The volume is filled with matter whose average comoving density is the mean background density
$\bar{\rho} = \Omega_{\mathrm{m}} \rho_{\mathrm{crit}} = 3\Omega_{\mathrm{m}} H_0^2/(8\pi G)$.
As mentioned before, we do not explicitly include baryons in our simulations.
In order to solve \cref{eq:SPscalar}, \code{AxiREPO} employs a second-order symmetrized split-step pseudo-spectral Fourier method.
For small time steps $\Delta t$, the time evolution can be approximated as follows \citep{Woo:2008nn}:
\begin{equation}
    \psi(t + \Delta t, \vec{x}) \approx
    e^{-i \frac{m}{\hbar} \frac{1}{a(t)} \frac{\Delta t}{2} \Phi(t + \Delta t, \vec{x})} \:
    e^{i \frac{\hbar}{m} \frac{1}{a(t)^2} \frac{\Delta t}{2} \nabla^2} \:
    e^{-i \frac{m}{\hbar} \frac{1}{a(t)} \frac{\Delta t}{2} \Phi(t, \vec{x})} \,
    \psi(t, \vec{x})
\end{equation}
where the time evolution operator has been split into three parts (\q{kick}, \q{drift}, \q{kick}) using the Baker–\allowbreak Campbell–\allowbreak Hausdorff formula.

The fields $\psi$ and $\Phi$ are discretized on a uniform Cartesian grid with $N$ grid points per dimension (for a total grid size of $N^3$), enabling numerical computations using the Fast Fourier Transform (FFT).
Accordingly, the numerical algorithm of the pseudo-spectral method performs a \q{kick} operation at the beginning and end of each time step,
\begin{equation}
    \label{eq:algorithm-kick}
    \psi \leftarrow e^{-i \frac{m}{\hbar} \frac{1}{a} \frac{\Delta t}{2} \Phi} \psi
    ,
\end{equation}
with a \q{drift} operation in between:
\begin{equation}
    \label{eq:algorithm-drift}
    \psi \leftarrow \FFT^{-1}\ml(e^{-i \frac{\hbar}{m} \frac{1}{a^2} \frac{\Delta t}{2} k^2} \FFT(\psi\mr)
    ,
\end{equation}
where $\FFT$ and $\FFT^{-1}$ denote the (Fast) Fourier transform and its inverse, respectively.
The gravitational potential $\Phi$ is similarly updated after each \q{drift} operation using a spectral solver for the Poisson equation:
\begin{equation}
    \Phi \leftarrow
    \FFT^{-1}\ml(
        \frac{1}{k^2} \FFT\ml(4\pi Gm\ml(|\psi|^2 - \langle|\psi|^2\rangle\mr)\mr)
    \mr)
\end{equation}
Consecutive executions of \cref{eq:algorithm-kick} (i.\,e.\ except for the initial and final time steps) can be combined into a single operation $\psi \leftarrow e^{-i \frac{m}{\hbar} \Delta t \Phi} \psi$ for efficiency.

The choice of the time step $\Delta t$ in \cref{eq:algorithm-kick,eq:algorithm-drift} follow from the requirement that the phase difference in the exponentials must not exceed $2\pi$ to prevent incorrect \q{aliasing} of the time step to a smaller value due to periodicity.
Both the kick (\cref{eq:algorithm-kick}) and drift (\cref{eq:algorithm-drift}) steps impose separate time step constraints, with the overall result:
\begin{equation}
    \label{eq:time-step}
    \Delta t < \min\ml(
        \frac{4}{3\pi} \frac{m}{\hbar} a^2 \Delta x^2,\;
        2\pi \frac{\hbar}{m} a \frac{1}{|\Phi|_{\mathrm{max}}}
    \mr),
\end{equation}
where $\Delta x = L / N$ is the spatial resolution and $|\Phi|_{\mathrm{max}}$ is the maximum (absolute) value of the potential.
The term involving the resolution $\Delta x$ is the drift constraint, while the term with the potential $\Phi_{\mathrm{max}}$ is the kick constraint.
The quadratic scaling of the time step with the spatial resolution, $\Delta t \propto \Delta x^2$, is typical for the Schrödinger–Poisson system, highlighting the relation of the Schrödinger equation to diffusion equations.

Beyond the time step, the velocity field also imposes a constraint on the spatial resolution.
Since the velocity is given by the \emph{gradient} of the complex field's phase (as can be seen from the continuity equation),
\begin{equation}
    \label{eq:velocity}
    \begin{aligned}
        \psi &= \sqrt{\frac{a^3 \rho}{m}} e^{i\theta}
        \\
        \bv &= \frac{\hbar}{m} \nabla \theta ,
    \end{aligned}
\end{equation}
whose difference between two points cannot exceed $2\pi$, the discretized velocity field has a maximum representable value (depending on the concrete form of the discretized gradient operator) of about
\begin{equation}
    \label{eq:velocity-criterion}
    v_{\mathrm{max}} = \frac{\hbar}{m} \frac{\pi}{\Delta x}.
\end{equation}
Velocities $v \ge v_{\mathrm{max}}$ cannot be represented in a simulation with resolution $\Delta x$.
This is equivalent to a spatial resolution constraint; $\Delta x$ should be small enough to resolve the de Broglie wavelength $\lambda_{\mathrm{dB}}(v) = 2\pi\hbar/(mv)$ of the largest velocities:
\begin{equation}
    \label{eq:velocity-criterion-resolution}
    \Delta x < \frac{\pi\hbar}{mv_{\mathrm{max}}} = \frac{1}{2} \lambda_{\mathrm{dB}}(v_{\mathrm{max}}).
\end{equation}

\begin{table}
    \centering
    \begin{tabular}{
        l c
        S[table-format=1.1]
        S[table-format=3.0]
        S[table-format=1.2e+2]
        S[table-format=1.3e1]
    }
        \toprule
        IC &
        $N^3$ &
        {$L$ / \si{\per\hHubble\Mpc}} &
        {$\Delta x$ / \si{\per\hHubble\pc}} &
        {$m$ / \si{\eV}} &
        {$k_*$ / \si{\per\Mpc}} 
        \\
        \midrule
        Iso.\ only & $1920^3$ & 0.4 & 208 & e-19 & e3 
        \\
        Iso.\ only & $1440^3$ & 0.4 & 278 & e-19 & e3 
        \\
        Iso.\ only & $\phantom{0}960^3$ & 0.4 & 417 & e-19 & e3 
        \\
        Iso.\ only & $2880^3$ & 0.8 & 278 & e-19 & e3 
        \\
        Iso.\ + adi.\ & $1920^3$ & 0.4 & 208 & e-19 & e3 
        \\
        Iso.\ + adi.\ & $2880^3$ & 0.8 & 278 & e-19 & e3 
        \\
        Iso.\ + adi.\ & $2160^3$ & 0.8 & 370 & e-19 & e3 
        \\
        Iso.\ + adi.\ & $1440^3$ & 0.8 & 556 & e-19 & e3 
        \\
        Iso.\ only, MB & $1920^3$ & 0.4 & 208 & 1.02e-19 & 0.455e3 
        \\
        \bottomrule
    \end{tabular}

    \caption{List of performed simulations with important characteristics: type of initial conditions (where most simulations used a uniform sphere for the initial field power spectrum, whereas ``MB" indicates a Maxwell–Boltzmann initial spectrum), number of grid cells $N^3$, box size $L$, resolution (grid cell size) $\Delta x$, boson particle mass $m$, and peak scale $k_*$.
    All simulations are evolved in time between $z_0 = \num{5e5} - 1$ ($a_0 = \num{0.006766} a_{\mathrm{eq}}$) and $z_{\mathrm{f}} = 99$ ($a_{\mathrm{f}} = \num{33.7} a_{\mathrm{eq}}$).
    For adiabatic initial conditions, $A_{\mathrm{s}} = \num{2.1412e-9}$ is used, and the value of the dimensionless power spectrum $(k^3/2\pi^2)P_\delta(k)$ at $k = \SI{1}{\per\Mpc}$ boosted to \num{e-4}.
    All lengths are comoving.}
    \label{tab:simulations}
\end{table}

The combination of the requirement to resolve structures on the scale of the de Broglie wavelength (\cref{eq:velocity-criterion-resolution}), and the time step requirement $\Delta t \propto \Delta x^2$ (\cref{eq:time-step}), results in much higher demands on computational resources for wave simulations compared to $N$-body simulations.

\subsubsection{List of Simulations}

\Cref{tab:simulations} shows the simulations from which we draw our results.
In addition to the cases listed there, we have run simulations at different particle masses $m$, but for the present work, we focus on a single value of $m = \SI{e-19}{\eV}$.

\subsubsection{Numerical Convergence}
\label{sec:convergence}

As shown in \cref{tab:simulations}, we performed identical simulations at different numerical resolutions to confirm that our results are numerically converged and do not suffer from resolution effects.
To demonstrate this, we compare the main quantities that make up our results in \cref{sec:SimResults} between different resolutions: The density power spectrum and the halo mass function.

\begin{figure}
    \centering

    \includegraphics{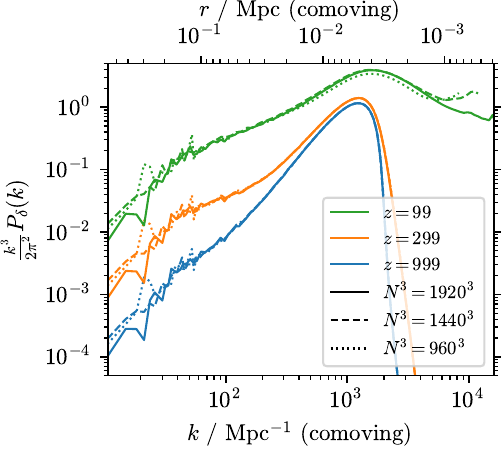}%
    \hfill%
    \hspace*{-0.8in}%
    \includegraphics{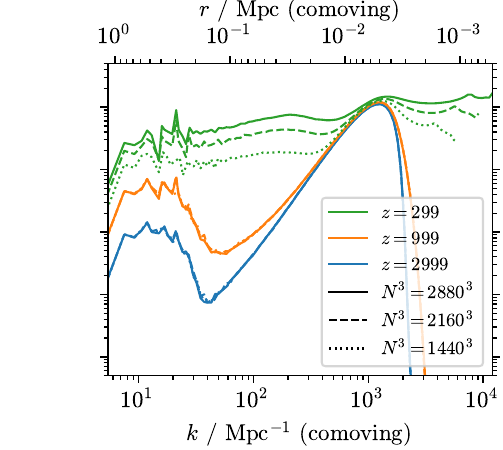}%

    \caption{%
        Comparison of the density power spectrum in simulations with different numerical resolutions, at different times.
        \textit{Left panel:} Simulations with isocurvature-only initial conditions ($L = \SI{0.4}{\per\hHubble\Mpc}$).
        \textit{Right panel:} Simulations with initial conditions including both adiabatic and isocurvature components ($L = \SI{0.8}{\per\hHubble\Mpc}$).%
    }
    \label{fig:convergence-powerspec}
\end{figure}

A comparison between the power spectra with different numerical resolutions is shown in \cref{fig:convergence-powerspec}.
For the isocurvature-only case, even at the comparatively low grid size of $N^3 = 960^3$, the evolution of the power spectrum agrees very well with the higher-resolution counterparts on almost all scales.
Only near the end of the simulation at $z = 99$ (where resolution requirements are most stringent) a slight global discrepancy appears between this lowest-resolution variant and the higher resolutions, demonstrating an excellent level of convergence for the highest resolution level of $1920^3$.

When including the adiabatic component of the power spectrum, a natural limit to the final simulation time results from the artificial enhancement of this component.
While this enhancement is necessary in order to capture a region where the adiabatic contribution is not negligible compared to the isocurvature one within the volumes accessible to simulations, it leads to an unphysical onset of non-linear structure formation on large scales at early times.
This effect is not only unphysical, but also spoils the ability of the simulation volume to be representative of a physical cosmic volume, since non-linear evolution occurs even on the largest scales, equal to the box size $L$.
For the value of the enhancement chosen here, this actually already occurs before the final simulation time of $z = 99$.

Correspondingly, we do not consider results beyond the onset of this unphysical non-linear evolution.
The right panel of \cref{fig:convergence-powerspec} shows that this is accompanied by a rapid increase in resolution requirements and significant resolution effects towards the end of the simulation.
However, since this only affects the unphysical regime, it is not a concern, and convergence before this time is once again excellent.

Generally, the largest differences appear on the smallest scales, at the resolution limit.
Not only are these scales at the edge of the spatial resolution, but they also correspond to the highest velocities (or equivalently, de Broglie wavelengths), which is most sensitive to (lack of sufficient) resolution.
In particular, the power spectrum is in the non-linear regime even at the smallest resolved scales, indicating that even higher resolution would be required to capture the details of the evolution on these extremely small scales.

A noteworthy, but not unexpected feature in the isocurvature case are the differences on the largest simulated scales, comparable to the size of the simulation box.
These are already present in the initial conditions and stay constant throughout the evolution.
Even though the initial field values are drawn using the same random number seed in all cases, the fact that the isocurvature initial conditions are imprinted on the complex field directly (rather than the density) results in random fluctuations in the density even with consistent random numbers.
These random fluctuations are the analog of cosmic variance and become visible on large scales because the number of modes decreases towards larger scales.

\begin{figure}
    \centering
    \includegraphics{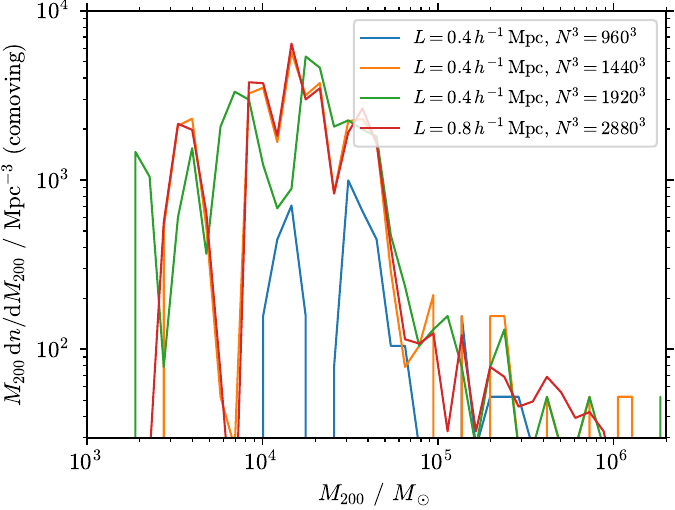}

    \caption{%
        Comparison of the halo mass function at the final time $z_{\mathrm{f}} = 99$ in simulations with different numerical resolutions.
        The simulations have isocurvature-only initial conditions; we did not study the HMF with adiabatic initial conditions since the required boost to the adiabatic power spectrum artificially leads to non-linear structure formation on large scales (in the adiabatic part) \emph{before} the isocurvature part.%
    }
    \label{fig:convergence-hmf}
\end{figure}

The behavior of the halo mass function at different resolutions is shown in \cref{fig:convergence-hmf}. While there is a clear difference between the lowest-resolution ($960^3$) simulation and higher resolutions, beyond a spatial resolution of $\Delta x \approx \SI{280}{\per\hHubble\pc}$ (corresponding to a $1440^3$ grid in the smaller box, or $2880^3$ in the larger box), there is not much difference. The exact locations of some peaks in the mass function vary slightly, but overall there is not a large difference between this intermediate resolution and the highest spatial resolution with $N^3 = 1920^3$ in the \SI{0.4}{\per\hHubble\Mpc} box.

Since these are simulations with isocurvature-only initial conditions, an interesting effect is that at fixed spatial resolution, the box size makes almost no difference for the shape of the mass function (see the orange and red lines in \cref{fig:convergence-powerspec}). This is because there is practically no power on large scales without the adiabatic contribution, so enlarging the box only adds large-scale mode without power which do not significantly contribute to structure formation. The main difference is the increased smoothness at the high-mass end, where objects are rarer. The improved statistics resulting from the eight-fold increase in simulation volume reduce fluctuations (noise) in this region where the halo mass function is low.

Given the persistent shape of the mass function with varying simulation volume and resolution (as well as different random realizations, not shown here), the observed irregularity of this shape is intriguing, although difficult to explain analytically (see also \cref{fig:plotMassFuncCompare}). A more detailed study of this mass function, as well as the optimal identification criteria for collapsed objects in this scenario, will be part of future work.

\end{document}